\documentclass{mn2e}
\usepackage{psfig}

\begin{document}

\title[CCD Photometry of GEHRs]{Narrowband CCD Photometry of Giant HII Regions}
\author[G. Bosch et al.]{Guillermo Bosch $^{1}$ \thanks{now at Facultad de 
Ciencias Astron\'omicas, La Plata, Argentina}, 
Elena Terlevich $^{2}$ \thanks{Visiting fellow IoA},
Roberto Terlevich $^{1}$ \thanks{Visiting Professor INAOE}
\\
$^{1}$ Institute of Astronomy, Madingley Road, Cambridge CB3 0HA \\
$^{2}$ INAOE, Puebla, M\'exico
}

\maketitle

\begin{abstract}
We have obtained accurate CCD narrow band $H\beta$ and $H\alpha$ photometry 
of Giant HII Regions in M~33, NGC~6822 and M~101.  Comparison with
previous determinations of emission line fluxes show large discrepancies;
their probable origins are discussed. Combining our new
photometric data with global velocity dispersion ($\sigma$) derived from 
emission line 
widths we have reviewed the  $L(H\beta)) - \sigma$ relation. A 
reanalysis of the properties of the GEHRs included in our sample
shows that age spread and the superposition of components in multiple 
regions introduce a considerable spread in the regression. Combining the
information available in the literature regarding ages of the associated 
clusters, evolutionary footprints on the interstellar medium, and
kinematical properties of the knots that build up the multiple GEHRs, we
have found that a subsample - which we refer to as young and single GEHRs -
do follow a tight relation in the $L - \sigma$ plane.

\end{abstract}


\section{Introduction}

Accurate photometry of  Giant HII Regions (GEHRs) is a key source
of information about the energetics of these large star-forming regions. The 
luminosity  emitted through the recombination lines of ionised
hydrogen (H$_{\alpha}$, H$_{\beta}$) allows to estimate
the number of ionising photons which in turn establish the requirements
for the stellar population responsible for such an ionising flux. 
Several reviews have already discussed these properties, such as
Kennicutt \shortcite{1979ApJ...228..394K},
Shields \shortcite{1990ARA&A..28..525S} and 
Melnick \shortcite{1992sfss.conf.....M}. 

These emission 
lines had  been found to have supersonic widths by Smith \& Weedman
\shortcite{1970ApJ...160...65S}, \shortcite{1971ApJ...169..271S}. 
Such behaviour
is puzzling, as supersonic motion cannot be sustained without
a source of energy to fuel it. Melnick \shortcite{1977ApJ...213...15M} 
and later 
Terlevich \& Melnick\ \shortcite{1981MNRAS.195..839T} 
found that the gas velocity dispersion ($\sigma$) measured from
a recombination line in  giant HII regions  correlates
well with the total luminosity ($L$) emitted in 
the respective line, 
and also with the overall radius of the nebula ($R$). 
In a logarithmic plane, total flux and nebula radius follow a 
linear relation with the line width \cite{1981MNRAS.195..839T} and the
following approximations hold:
$L \propto \sigma^4$ and $R \propto \sigma^2$.
These dependences resemble much those of
virialised systems, such as globular clusters and elliptical galaxies,
and lead the authors to suggest a gravitational explanation for
this observed kinematical behaviour.
Other authors confirmed
the existence of a correlation, although they found different slopes
for the linear regression. New models were produced to 
account for the underlying dynamics. Turbulence 
($L \propto \sigma^3$, Hippelein \shortcite{1986A&A...160..374H}) and multiple shells
($L \propto \sigma^6$, Arsenault \& Roy \shortcite{1988A&A...201..199A}) were introduced 
in the models. 
Fuentes Masip et al.\ \shortcite{2000AJ....120..752F} have obtained imaging
spectroscopy of the irregular galaxy NGC~4449 with which they detected and
analysed its giant HII regions. They studied the correlation of size
and luminosity with the velocity dispersion and found a regression for
the subsample that showed supersonic motions, high surface brightness and
single Gaussian emission lines. This linear regression in the log plane had
a slope of $3.8 \pm 1.2$ which is consistent with a gravitational origin of
the supersonic motions.

The intriguing aspect of the $L$--$\sigma$ relation is that although we know 
that massive stars produce strong winds and in many cases we can see features 
that {\it look like} shells, the scaling laws seem to suggest that gravity may 
play a central role.
Attempts to determine the main broadening mechanism in HII regions with 
supersonic linewidths were made for the 
closest and most luminous GEHRs within the Local Group, 30 Doradus in the
Large Magellanic Cloud and NGC~604 in M~33. 
Chu \& Kennicutt \shortcite{1994ApJ...425..720C} used longslit
echelle spectroscopy data on several positions of 30 Doradus and found that its
global supersonic linewidth breaks up into a large number of
smaller supersonic structures, which combine with a smooth underlying
turbulent velocity field. They suggested that gravitation could not 
contribute significantly to the observed nebular kinematics. This
finding was partly confirmed by Melnick, Tenorio-Tagle and Terlevich
\shortcite{1999MNRAS.302..677M} who obtained a single longslit echelle
spectrum with higher resolution and signal to noise ratio. The slit used
to obtain the spectrum, passed through several important features which are
analysed in their paper. However, they find no discrete component with
supersonic velocity widths, beyond a very broad component ($\sigma \sim$
45 km s$^{-1}$) found towards all lines of sight. This does not seem to be 
the case for NGC~604. Mu\~noz-Tu\~n\'on et~al. \shortcite{1996AJ....112.1636M},
analysed high spatial and spectral resolution H$\alpha$ emission line data
on NGC~604 
and concluded that although there is a 
contribution from
shells and filaments to the observed gas kinematics, the overall
supersonic behaviour is dominated by unresolved components, observed
throughout the whole intensity range. By modelling the expected behaviour
of the emission line supersonic width and intensity of a GEHR 
they suggested evolutionary effects to explain their different observed
behaviour. In a similar study, Yang et al.~\shortcite{1996AJ....112..146Y}
concluded that a substantial fraction of the broadening must be due to 
gravity.

The regression among global characteristics of GEHRs also holds for HII 
Galaxies. These are luminous star forming regions that dominate the galaxian 
energy output.  The relation between a distance independent parameter, such as 
the linewidth,
and the luminosity of the object allows us to use them as distance
indicators. The strength of the flux emitted in the recombination lines
of these objects makes them available as far as $z \simeq 3$, and even 
further with the new generation large telescopes. 
Lowenthal et al.\ \shortcite{1997ApJ...481..673L} have already obtained 
Keck spectroscopy of a sample of $z \sim 3$ galaxies from the Hubble Deep 
Field project
\cite{1996AJ....112.1335W}. They found that these objects seem to lie at 
the bright end of the luminosity - size relation present for HII galaxies 
although the starburst origin of their observed spectra remains to be
proven. Melnick et al.\ \shortcite{2000MNRAS.311..629M},
making use of Guzm\'an et al.\ \shortcite{1997ApJ...489..559G}
together with Koo et al.\ \shortcite{1996ApJ...469..535K} and Pettini et al.\
\shortcite{1998ApJ...508..539P} data for high $z$,  have 
re-analysed the luminosity - linewidth regression for HII galaxies and 
found a good correlation up to z $\sim$ 3, in spite of several 
systematic effects that might be increasing the observed scatter in the
regression.

Previous studies dealing with the $ L - \sigma $ regression for
the Giant HII regions have used 
datasets from a variety of sources and detectors, which may lead to large
internal errors of the samples. The largest and more homogeneous 
available photometric data in the literature for the H$\alpha$ 
emission line in GEHRs is K79,
which is used by Hippelein \shortcite{1986A&A...160..374H} and 
Arsenault \& Roy \shortcite{1988A&A...201..199A} in their studies.
H$_{\beta}$ photometry was 
performed by Melnick \shortcite{1979ApJ...228..112M}
and later by Melnick et al.\ \shortcite{1987MNRAS.226..849M}. 
These data are discrepant, reasons for that could be the use of non-linear 
detectors, the lack of good standards for narrowband emission line photometry, 
or uncertainties in the  extinction corrections.

The disagreements between previous luminosity estimations of GEHRs suggests
that errors in the determination of fluxes may be a main contributor
to the observed spread in the  $ L - \sigma $ regression. 
The goal of the present
photometric study is to achieve a standard set of H$\alpha$ emission 
fluxes from nearby extragalactic giant HII regions to use as the `local'
reference system. 
To this end we have performed CCD photometry on a set
of GEHRs from galaxies with relatively well known distances, in order to
calculate their absolute fluxes. The observations are described in Section 
\ref{sec:photobs},
the data reduction is described in Section \ref{sec:photdatred}, and 
Section \ref{sec:photfluxcal} introduces the
flux calibration process designed for this work. The procedures used to obtain
the total fluxes in each band are listed in Section \ref{sec:alphabeta} 
and the results, plus comparison with previous studies are 
shown in Section \ref{sec:photresults}.

Extinction maps are presented in Section \ref{sec:extinction}.
The relation between the luminosity and width of the emission lines is 
discussed in Section \ref{sec:lsigma} 
and conclusions are drawn in Section \ref{photdiscussion}.

\section{Observations}
\label{sec:photobs}

\begin{figure}
\centerline{
\fbox{Figure1}}
\caption{Finding chart for NGC~ 588. The axes show the angular scale, estimated
from converting the pixel to arcsec scale for the TeK CCD at the JKT focal 
plane. The coordinates zero point coincides with the origin of the curve of 
growth method used to find the total flux of the region.}
\label{fig:ngc588}
\end{figure}
\begin{figure}
\centerline{
\fbox{Figure2}}
\caption{Same as figure \ref{fig:ngc588} for NGC 592.}
\label{fig:ngc592}
\end{figure}
\begin{figure}
\centerline{
\fbox{Figure3}}
\caption{Same as figure \ref{fig:ngc588} for NGC 595.}
\label{fig:ngc595}
\end{figure}
\begin{figure}
\centerline{
\fbox{Figure4}}
\caption{Same as figure \ref{fig:ngc588} for NGC 604.}
\label{fig:ngc604}
\end{figure}
\begin{figure}
\centerline{
\fbox{Figure5}}
\caption{Same as figure \ref{fig:ngc588} for NGC 5447.}
\label{fig:ngc5447}
\end{figure}
\begin{figure}
\centerline{
\fbox{Figure6}}
\caption{Same as figure \ref{fig:ngc588} for NGC 5461.}
\label{fig:ngc5461}
\end{figure}
\begin{figure}
\centerline{
\fbox{Figure7}}
\caption{Same as figure \ref{fig:ngc588} for NGC 5462.}
\label{fig:ngc5462}
\end{figure}
\begin{figure}
\centerline{
\fbox{Figure8}}
\caption{Same as figure \ref{fig:ngc588} for NGC 5471.}
\label{fig:ngc5471}
\end{figure}
\begin{figure}
\centerline{
\fbox{Figure9}}
\caption{Same as figure \ref{fig:ngc588} for NGC 6822~I.}
\label{fig:ngc6822I}
\end{figure}
\begin{figure}
\centerline{
\fbox{Figure10}}
\caption{Same as figure \ref{fig:ngc588} for NGC 6822~III.}
\label{fig:ngc6822III}
\end{figure}

CCD frames of the selected Giant HII Regions were obtained during seven nights
at the 1.0m Jacobus Kapteyn Telescope (JKT) in the Observatorio del Roque de 
los Muchachos (ORM) La Palma, Spain, in August 1997. Five out of the seven 
nights
were of excellent photometric quality; a sixth one, although not photometric, 
was useful too. The last
night of the run had variable sky conditions so the data was not included in
the analysis. The average seeing conditions were monitored from the image 
quality of short exposures of the flux standard stars, and an upper limit
was also estimated from stellar sources present in our longer exposed
nebular frames. These estimations, and the atmospheric extinction measurements 
available from the Carlsberg Meridian Circle, also located at La Palma,
 are listed in Table~\ref{seeing}.

\begin{table}
\begin{center}
\begin{tabular}{|l|cc|}
\hline
Date & Seeing & Extinction \\
     & (average) & (mag.) \\
\hline
09/08/97 & 0\farcs90  & .11   \\
10/08/97 & 0\farcs86  & .11   \\
11/08/97 & 0\farcs91  & n/a   \\
12/08/97 & 1\farcs02  & .11   \\
13/08/97 & 1\farcs00  & .11   \\
14/08/97 & 0\farcs96  & .19   \\
15/08/97 & $\geq 1\farcs12$ & .18   \\
\hline
\end{tabular}
\end{center}
\caption[Sky conditions for the La Palma sky during the observing run]{Sky 
conditions for the La Palma sky during the observing run. Seeing
conditions were measured from stellar sources on the CCD images. The extinction
data is provided by the Carlsberg Meridian Circle (La Palma)}
\label{seeing}
\end{table}

The target objects, together with the flux standard stars were observed using
four narrow-band filters. These are centred in H$\alpha$ (6561\AA), H$\beta$ 
(4861\AA), 
and their adjacent continua (6656\AA\ and 5300\AA\ respectively). 
The main characteristics of the 
set of filters used are given in Table~\ref{filters}. Three of them are from 
the ING bank of filters. The H$\beta$ continuum filter
was kindly provided by Don Pollaco, as the 
presence of nebular emission lines, such as [OIII]$\lambda\lambda$ 4959,5007
\AA~made the available continuum filters useless. 

\begin{table}
\begin{center}
\begin{tabular}{|l|cccc|}
\hline
Filter & ING\# & $\lambda_c$ & $\Delta\lambda$ & Peak Tr. \\
\hline
H$\alpha$        & 60 & 6561 & 38 & 33.9  \\
H$\alpha_{cont}$ & 67 & 6656 & 44 & 47.5 \\
H$\beta$         & 81 & 4861 & 54 & 59.7 \\
H$\beta_{cont}$  & \ldots & 5300 & 50 & 58.5 \\
\hline
\end{tabular}
\end{center}
\caption[Description of the filters used.]{Description of the filters used. Column 2 indicates the 
corresponding number in the Isaac Newton Group (ING) database. Column 3 lists
the central wavelength of each filter (in \AA ), 
column 4 the bandpass width (in \AA ), and column 5 their peak transmission
values (in percentage)}
\label{filters}
\end{table}

The target objects from three different  Local Group galaxies, namely 
NGC~6822, M~33, and M~101, were widely spread in right ascension. This allowed
to sequentially observe them quite close to their passage through the local
meridian, reducing the effects of large airmass corrections. On the other hand,
the objects were observed on different days at different bandwidths, which
complicates the  calibration.
Table~\ref{logobsjkt} gives the observing log and figures \ref{fig:ngc588}
to \ref{fig:ngc6822III} show the emission line flux distribution in the sky.
These figures include a scale bar, as a guideline for linear sizes of the 
regions analysed in this paper.

\begin{table}
\begin{center}
\begin{tabular}{|l|p{1.35cm}p{1.4cm}p{1cm}p{1cm}|}
\hline
Object & H$\alpha$ & H$\alpha_{cont}$ & H$\beta$ & H$\beta_{cont}$ \\
\hline
NGC~588 & 12 & 12 & 14 & 14 \\
NGC~592 & 11 & 11 & 15 & 15 \\
NGC~595 & 10 & 10 & 10 & 14 \\
NGC~604 &  9 &  9 & 13 & 13 \\
NGC~5447 & 13 & 13 & \ldots & \ldots \\
NGC~5455 & 15 & 14 & \ldots & \ldots \\
NGC~5461 & 11 & 11 & \ldots & \ldots \\
NGC~5462 & 11 & 11 & \ldots & \ldots \\
NGC~5471 & 10 & 10 & \ldots & \ldots \\
NGC~6822I & 9  &  9 & 10 13 & 11 14 \\
NGC~6822III & 9  &  9 & 10 13 & 11 14 \\
BD+$28^{\circ}4211$ & 9$^{\star}$ 10 11$^{\star}$ 12$^{\star}$ 13 15$^{\star}$
& 9$^{\star}$ 10 11$^{\star}$ 12 13 14$^{\star}$ 
& 10 13 14$^{\star}$ 15$^{\star}$
& 11$^{\star}$ 13 14$^{\star}$ 15$^{\star}$ \\
BD+$33^{\circ}2642$ & 10$^{\star}$ 11$^{\star}$ 12 13$^{\star}$ 15  &
 10$^{\star}$ 11$^{\star}$ 12 13 14 &
10$^{\star}$13$^{\star}$ 14 15  &
11$^{\star}$  13$^{\star}$ 14 15  \\
Feige110 & 9 & 9 & \ldots & \ldots \\
\hline
\end{tabular}
\end{center}
\caption[Observing log for the target and calibration objects.]{Observing log for the target and calibration objects. The number
represents the date in which the object was observed. For the standard stars,
a `$^{\star}$' indicates more than one observation in that day.}
\label{logobsjkt}
\end{table}
Bias frames and evening and morning flats were taken every day. The importance
of the double set of flats is discussed in the following section.

\section{Data Reduction}
\label{sec:photdatred}

Standard procedures using IRAF software were followed for the initial steps
of the data reduction.
Bias frames were combined and used for calculating the
additive zero point correction as usual. The correction for flat field, 
however, was more delicate as the 
nature of the flux calibration process discussed in the next section relies
strongly on the flatness of the field illumination. 

As mentioned above, we obtained blank sky exposures to correct from
non-uniform illumination of the CCD\@. A simple check  consisting on
correcting one set of
flat field images, say the evening ones, with the ones obtained the morning
after, showed a non-uniform image with patterns still present in it.
Figure~\ref{2dflakes} shows a section of a flatfield image taken with
the H$\alpha$  filter. Two sets of annular patterns can be seen. Some are 
small with clearly defined edges while  others are larger with difuse edges.
The smaller rings are neatly removed but the diffuse annuli remain. Ring-like
patterns are usually associated with grains of dust, the reason for their
appearance being that they are observed severely out of focus; 
they are usually referred to as `dust flakes'.
The observed displacement might be explained if large rings
are present in the filters. As the filter 
wheel does not park exactly in the same position each time the same
filter is requested, the ring patterns are shifted. The smaller rings might
therefore be present on other surfaces (i.e. dewar window) 
The dust flakes problem has its maximum
effect in the H$\alpha$ filter, and is negligible for the other ones.

\noindent
\begin{figure}
\centerline{
\fbox{Figure11}}
\caption[Section of the evening H$_{\alpha}$ flatfield frame.]{Section of the evening H$_{\alpha}$ flatfield frame. Note the large
ring structure throughout the whole region, and the smaller, sharper rings.}
\label{2dflakes}
\end{figure}

\section{Flux calibration}
\label{sec:photfluxcal}

\subsection{Standard star magnitudes}

After the preliminary steps mentioned in the previous section, the next
stage is the flux calibration of our photometry. 
Narrowband emission line photometry is unusual. There
is no set of standard stars calibrated as to perform a routine
reduction process. There is no analog to the UBV system to use
with H$\alpha$, H$\beta$ and their respective continuum filters.
So to obtain photometry of emission line regions,
like  the GEHRs, some ad-hoc calibration has to be adopted. To achieve this, 
we have chosen three  HST spectrophotometric stars (BD+$28^{\circ}4211$, 
BD+$33^{\circ}2642$, and Feige110 \cite{1990AJ.....99.1243T}) 
for which 
flux is calibrated and tabulated every 1 or 2 \AA\@. This high resolution
allows to calculate their absolute flux (or magnitude) through every
single filter of our system. This can be achieved integrating numerically,
as the 
transmission curves of the used filters, scanned at the ORM, are available. 
We can define the absolute flux in
a filter as:
\begin{equation}
 F_{*}(\lambda_0) = \frac{\int{F_{*}(\lambda) \, T(\lambda) d\lambda}}
{\int{T(\lambda) d\lambda}} 
\label{eq:filterflux}
\end{equation}
where $\lambda_0$ is the central wavelength of the filter with transmission
curve $T(\lambda)$ and $F_{*}(\lambda)$ is the absolute flux of the star.

The calibration magnitude of the observed standard
stars for each filter was calculated as $m=-2.5 \, \log F$,
as it would be measured outside the earth's atmosphere.
The results are listed in Table~\ref{stdmags}.

\begin{table}
\begin{center}
\begin{tabular}{|l|r|r|r|}
\hline\hline
                   & BD~28$^\circ$4211 & BD~33$^\circ$2642 & Feige~110 \\
\hline
$F(H_{\alpha})$    & 1041.30 & 922.61 & 304.23 \\
$mag(H_{\alpha})$  & -7.5439 & -7.4125 & -6.2080 \\
\hline
$F($6650\AA)        & 1070.84 & 881.30 & 320.41 \\
$mag($6650\AA)      & -7.5743 & -7.3628 & -6.2643 \\
\hline
$F($5300\AA)        & 2575.10 & 1901.85 & 764.53 \\
$mag($5300\AA)      & -8.5270 & -8.1979 & -7.2085 \\
\hline
$F(H_{\beta})$     & 3484.94 & 2475.61 & 995.02 \\
$mag(H_{\beta})$   & -8.8555 & -8.4842 & -7.4946 \\
\hline\hline
\end{tabular}
\caption{Calculated fluxes and magnitudes for the standard stars used
for the calibration of the photometry.}
\label{stdmags}
\end{center}
\end{table}

\subsection{Atmospheric extinction correction}

The disadvantage of working with these flux standard stars is the limited 
number of observations per night, as there is 
only one star per field. 
Nevertheless, as  we had perfect photometric conditions during the first
four nights of our run,  we could gather the data available from these nights
to perform the extinction correction. 
Also, within 
our own photometric system, the transformation from observed to calibrated
magnitudes does not require the introduction of colour corrections. The 
transformation equations are written as
\begin{equation}
m_{calib} = m_{obs} + Z_{\lambda} + k \, X 
\label{eq:calmag}
\end{equation}
where $m_{calib}$ is the star calibrated magnitude and 
$m_{obs}$ is the observed
magnitude, $Z_{\lambda}$ is the zero-point correction for each filter (to 
account for slight differences between the tabulated and real transmission 
of the filter), $X$ is the normalised airmass ($X=1$ at zenith) and $k$ is
the extinction in magnitudes due to the airmass $X$.

\begin{figure}
\centerline{
\psfig{figure=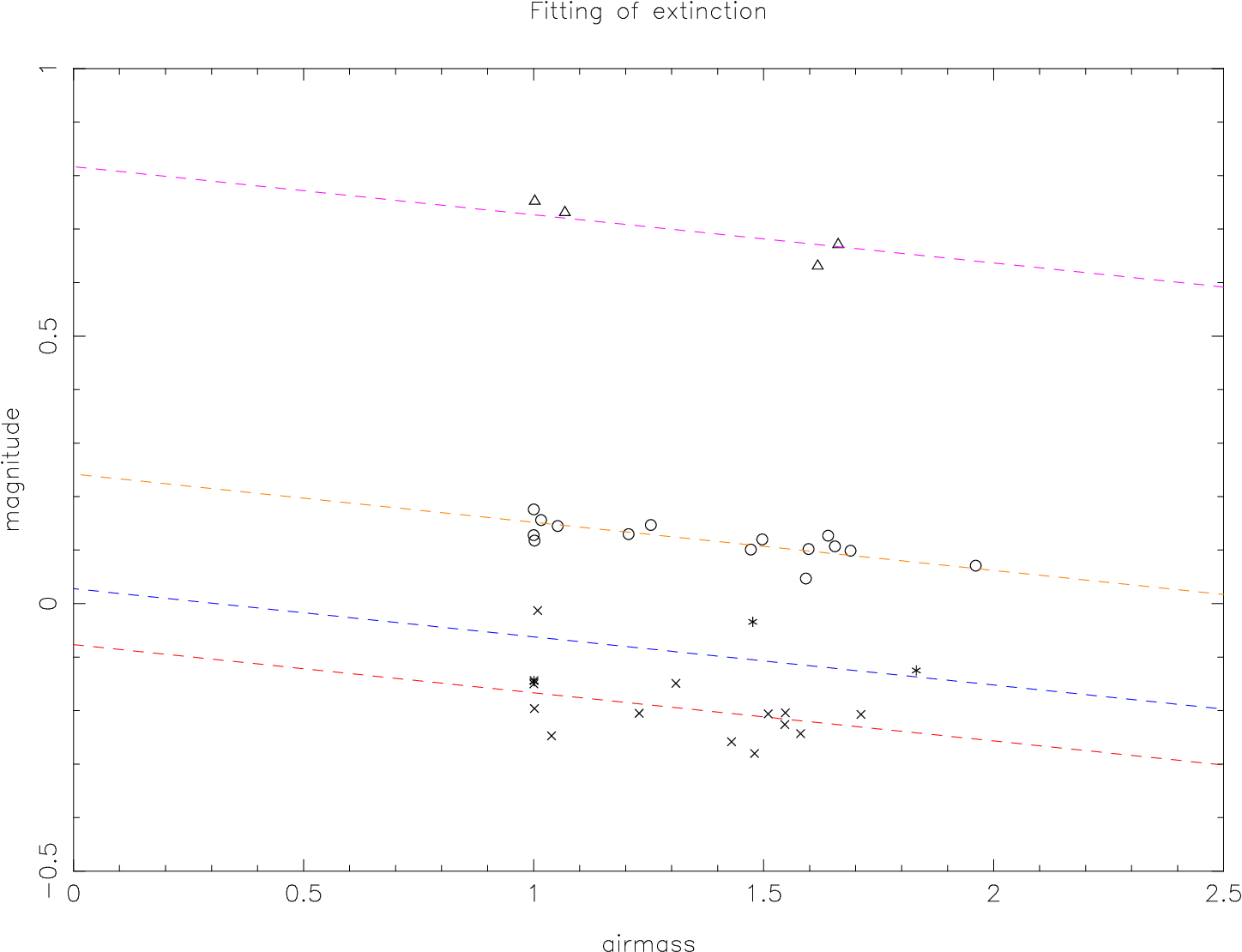,angle=270,width=8cm,height=8cm,clip=}}
\caption[Visualisation of the result of the least squares fit to the 
extinction correction.]{Visualisation of the result of the least squares fit 
to the
extinction correction. Each filter is represented by a different symbol: 
open triangles indicate the 5300 \AA~filter, open circles the
6650 \AA~filter, asterisks the H$\beta$ filter, and crosses show data for
the H$\alpha$ filter. Different stars are included in each filter group. 
The point where the extrapolation of the lines drawn meet the $X=0$ axis
indicate the zero-point correction. Within the observational errors the 
assumption of a 
single slope for all filters is valid.}
\label{extinctfig}
\end{figure}

The magnitudes of the standard stars were obtained performing aperture
photometry on the observed stars, using IRAF's version of DAOPHOT\@.
We considered a different value of Z for each filter, but assumed the same
value of $k$ for all filters. As we are working within a small range of
wavelengths (4860 - 6650 \AA)  this is a reasonable approach.
The set is solved numerically by least squares.
The results are shown in Figure~\ref{extinctfig} and listed 
in Table~\ref{extincttab}.

\begin{table}
\begin{center}
\begin{tabular}{|l|ll|}
\hline\hline
Parameter        & Fit Value & Error \\
\hline
$Z_{H_{\alpha}}$ & -0.0766 & 0.036 \\
$Z_{6650}$       & ~0.2421 & 0.037 \\
$Z_{H_{\beta}}$  & ~0.0279 & 0.046 \\
$Z_{5300}$       & ~0.8166 & 0.042 \\
$k$              & -0.0900 & 0.026 \\
\hline
\end{tabular}
\caption[Zero point correction and extinction per airmass unit values]{Values 
from the least squares fit to solve for the
extinction correction. $Z_{\lambda}$ indicates the zero-point correction
and k is the extinction per airmass unit. All quantities are
in magnitudes.}
\label{extincttab}
\end{center}
\end{table}

The set of defined {\bf extinction corrected magnitudes} is
applied  to the following nights, when the sky conditions are expected
to be slightly different. For these nights we can  apply the results
of the calibration, where the $k$ parameter is known, and  find
the new value of $Z_{\lambda}$, which might differ from the previous one
due to atmospheric changes. The values of $\Delta Z_{\lambda}$ were
checked, in order to see if conditions also changed during the night,
which was the case for the last night
of observations. The number of standard stars observations were not enough
to quantify this variability and account for it, so we decided not to use the
data from the last night.

\subsection{Calibrating the frames}

Having flatfielded the frames and defined a
relation between absolute flux (or magnitude) and the number of counts
detected, we can use the 
flux-to-counts relation established for the standard stars throughout the 
whole field of the CCD and apply it to our program objects. Before this, 
images were aligned, cosmic ray cleaned and had their continuum contribution 
subtracted following standard procedures. More details can be found in
Bosch \shortcite{1999..Bosch..PhDTh..CamUK}.

The flux standard stars have almost featureless continua. Hence the flux 
remains almost constant throughout the whole filter range. This is not the
case for the line flux emitted by the nebulae, as the emission is concentrated
in less than one \AA .  The factor by which we are
 underestimating the line flux when assuming it has a flat distribution is 
calculated. The
true flux can be obtained making use of spectroscopy already published 
in the literature from which we know the central wavelength and
dispersion of the emission line for our objects. Therefore we can construct a 
Gaussian 
emission line
\begin{equation}
\frac{1}{\sqrt{2\pi}\sigma} \exp(-\frac{(\lambda - \lambda_0)^2}{2 \sigma^2})
\label{eq:gaussian}
\end{equation}
where $\lambda_0$ and $\sigma$ are selected according to the known
radial velocities and observed line widths.
Galaxies in the Local Group do not have large radial velocities, so the
wavelength of the emission lines are close to the rest value. The 
typical width ($\sigma$) of the emission lines comes from their supersonic
velocity dispersion, $\sim$ 30 km s$^{-1}$, which corresponds to a width of 
approximately 0.67~\AA~for the H$\alpha$ line.

The Gaussian emission line defined in Eq.\ref{eq:gaussian} has unit flux 
by definition, so we can use a 
similar procedure to the one used for the continuum. In this case we have to
calculate the throughput of this artificial line through the filter,
\begin{equation}
F_{neb}(\lambda_0) = \frac{\int{F_{neb}(\lambda) \, T(\lambda) d\lambda}}
{\int{T(\lambda) d\lambda}},
\end{equation}
which yields the conversion factor,
$(F_{neb}(\lambda_0))^{-1}$, necessary to correct from the observed flux into
true flux for the relevant emission line.

The narrow $H\alpha$ emission line, is not the only contribution to the 
measured flux in the $H\alpha$ filter, due to the presence of the forbidden 
[NII] $\lambda\lambda 6548,6584$ \AA\ lines within the range of the filter.
In order to estimate -- and correct for -- the intensity of these lines, we 
have gathered information available in the literature about their
relative intensities with respect to $H\alpha$ and introduced
in the simulation described above, Gaussian profiles with their
respective intensities. The net expected true intensity from the
$H\alpha$ emission has then been calculated.

The line intensities from the literature are listed in Table~\ref{nitrogen2}.
In the case of NGC~5462 and NGC~5447 only the total value for both lines was 
available in Smith \shortcite{1975ApJ...199..591S}. 
To follow the procedure described above, we need to introduce intensities for
each [NII] line in the simulated spectrum in order to calculate their 
relative effect due to the observed redshift. The intensity for each of the 
[NII] lines was obtained from the total one, considering 
average intensity ratios 
between both [NII] lines ($\lambda 6548/\lambda 6584 \sim 3$). The value for 
NGC~592 (which doesn't exist in the literature) was estimated considering
the fact that there is a radial gradient in the abundance of nitrogen
detected in M~33 \cite{1988MNRAS.235..633V}
and that the region is 3.4 kpc away from the centre of the galaxy, quite close
to the radial distance of NGC~604 and NGC~595. An average of the intensities of
the [NII] lines for
these two regions was used to estimate the intensity of the lines 
in NGC~592.

\begin{table}
\begin{center}
\begin{tabular}{lcccc}
\hline\hline
Region & [NII] $\lambda 6548$ & [NII] $\lambda 6584$ & 
$k$ & Ref. \\
\hline
NGC 588 & 0.012 & 0.034 & 0.975 & 1 \\
NGC 592 & 0.045 & 0.125 & 0.912 & * \\
NGC 595 & 0.047 & 0.134 & 0.903 & 2 \\
NGC 604 & 0.043 & 0.117 & 0.907 & 1 \\
NGC 5447 & 0.023 & 0.070 & 0.961 & 3 \\
NGC 5461 & 0.037 & 0.110 & 0.941 & 4 \\
NGC 5462 & 0.040 & 0.119 & 0.937 & 3 \\
NGC 5471 & 0.008 & 0.026 & 0.987 & 4 \\
NGC 6822I & \ldots & 0.018 & 0.992 & 5 \\
NGC 6822III & \ldots & 0.019 & 0.992 & 5 \\
\hline
\end{tabular}
\caption[NII intensities and correction factors]{Intensities for the [NII] 
$\lambda\lambda 6548,6584$ \AA\ 
lines, relative to H$\alpha$, and the correction factor derived. 
Reference values are from: (1) V\'\i lchez et al.\ 
\shortcite{1988MNRAS.235..633V},
(2) Kwitter \& Aller \shortcite{1981MNRAS.195..939K}, 
(3) Smith \shortcite{1975ApJ...199..591S}, 
(4) Rayo et al.\ \shortcite{1982ApJ...255....1R},
(5) Pagel et al.\ \shortcite{1980MNRAS.193..219P}, 
(*) Averaged from the intensities of NGC 595 \& NGC 604 (see text).}
\label{nitrogen2}
\end{center}
\end{table}

\section{$H\alpha$ and $H\beta$ photometry}
\label{sec:alphabeta}

\begin{figure}
\centerline{
\psfig{figure=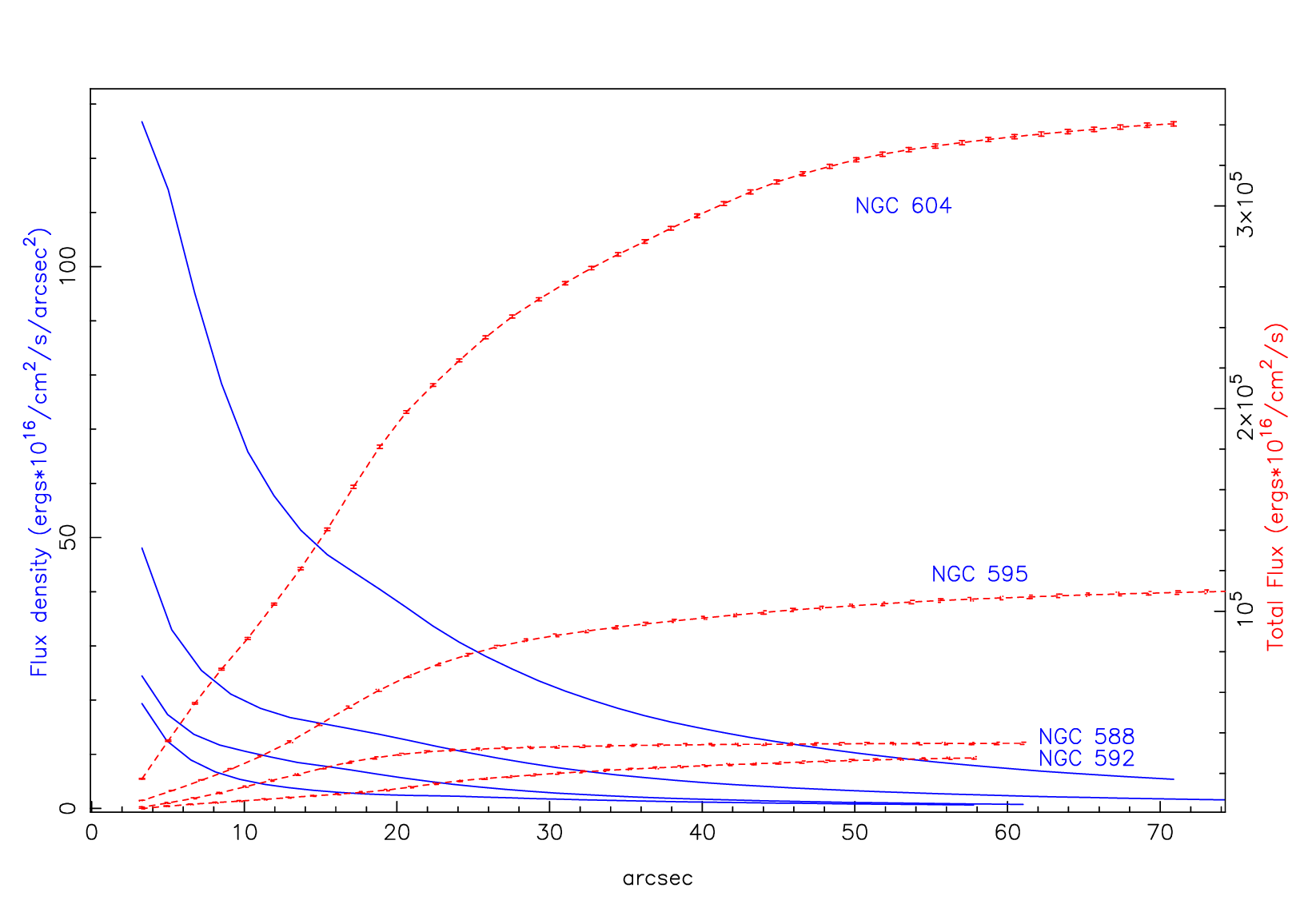,angle=270,width=8cm,height=8cm,clip=}}
\caption{Radial and cumulative flux density for the GEHRs in M~33. 
X-axis indicate radial distance to the region centre. Radial flux
density profiles (solid lines) are labelled in the left y-axis, The incremental
total flux profile (dashed lines) are labelled in the right y-axis. Each
region is identified on the cumulative flux plot. The top to bottom order
holds for the radial density plots.}
\label{fig:M33plot}
\end{figure}

\begin{figure}
\centerline{
\psfig{figure=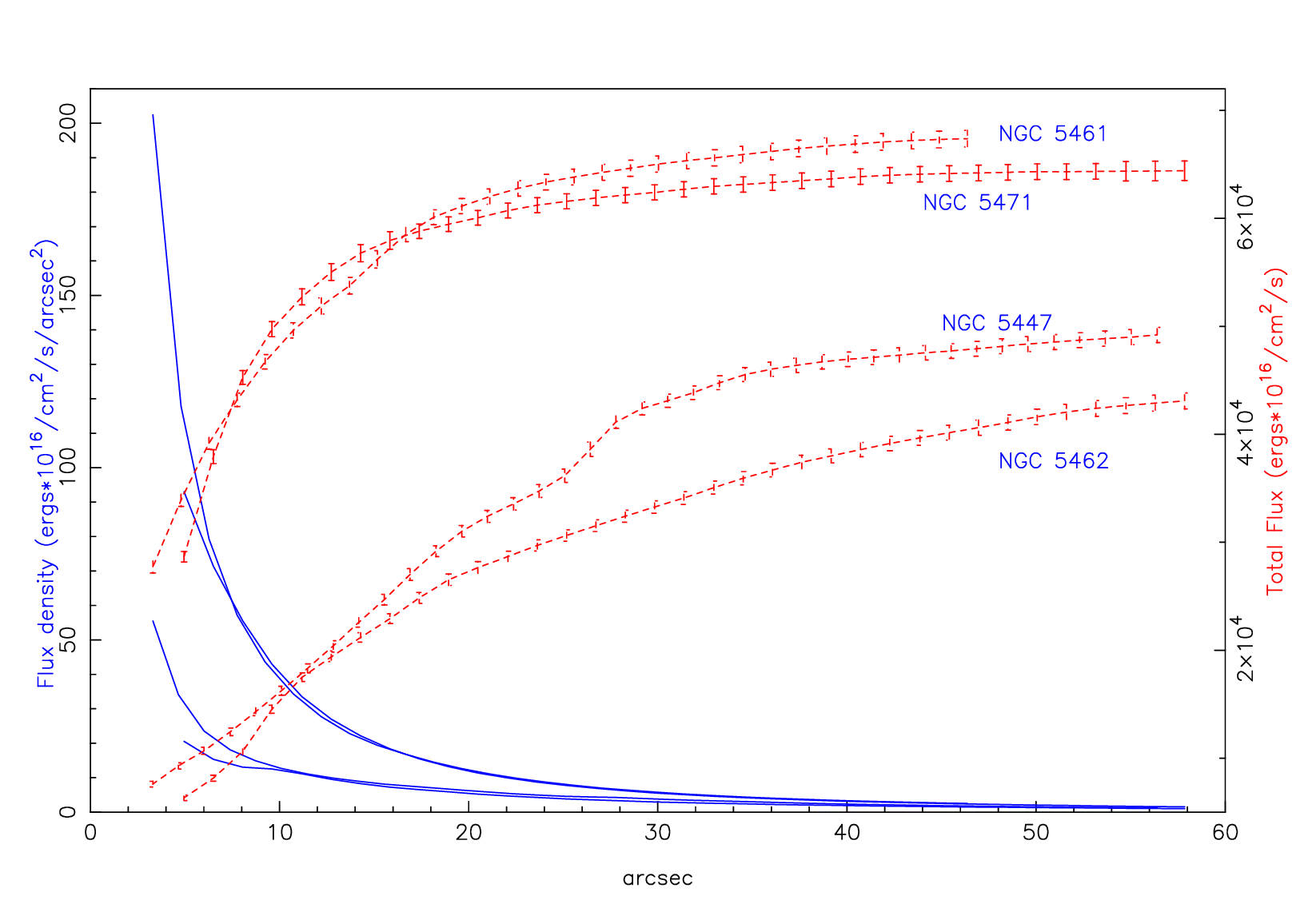,angle=270,width=8cm,height=8cm,clip=}}
\caption{Radial and cumulative flux density for the GEHRs in M~101. Axes
and labels are as in figure \ref{fig:M33plot}.}
\label{fig:M101plot}
\end{figure}

\begin{figure}
\centerline{
\psfig{figure=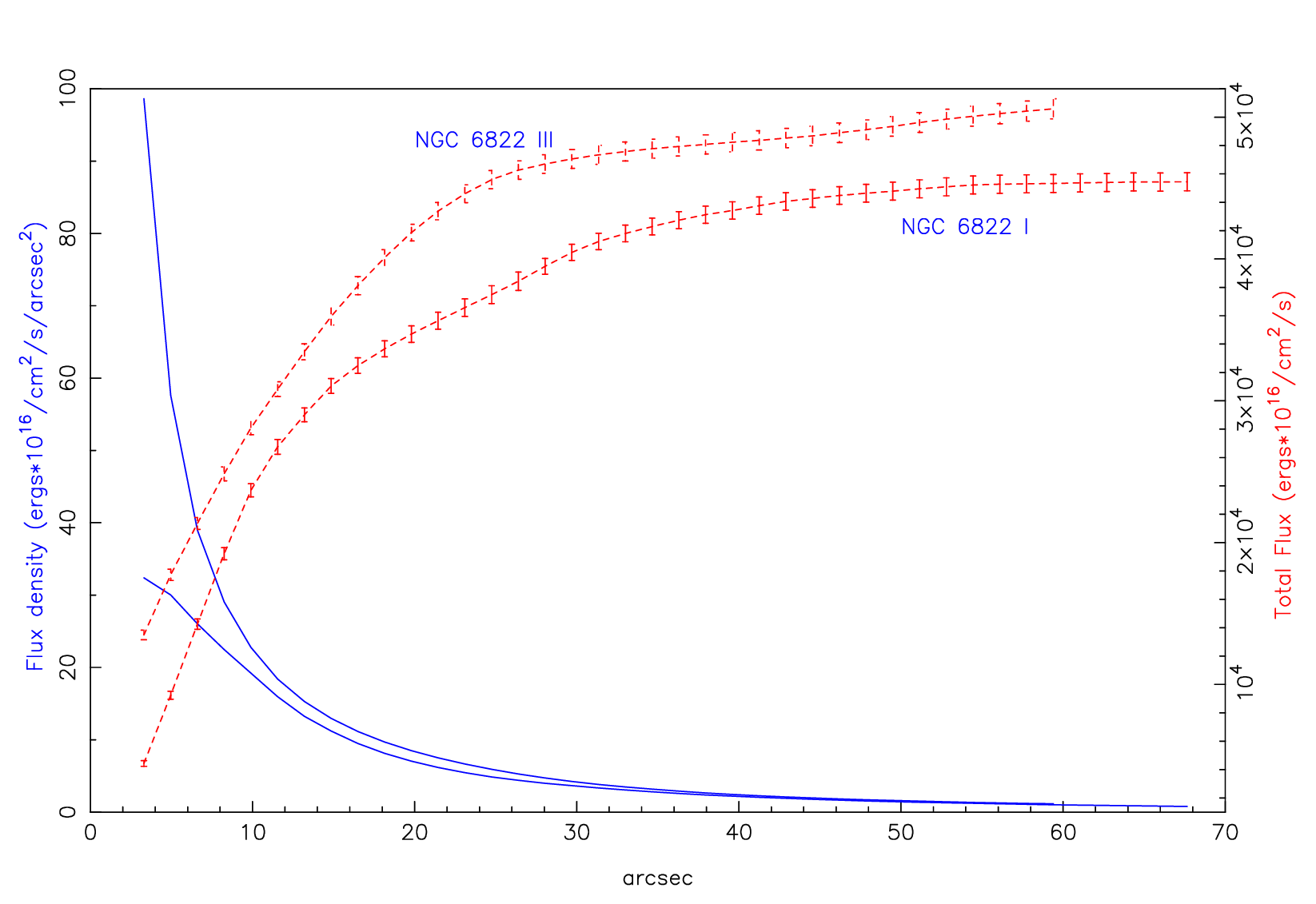,angle=270,width=8cm,height=8cm,clip=}}
\caption{Radial and cumulative flux density for the GEHRs in NGC 6822. Axes
and labels as in figure \ref{fig:M33plot}.}
\label{fig:NGC6822plot}
\end{figure}

The photometric analysis was done with the {\sc PHOTOM}, {\sc GAIA} and 
{\sc ESP} packages from {\sc STARLINK}. These allow to measure flux within 
increasing radius from the centre
and build a curve of growth in order to obtain the total flux of the GEHR 
in each emission line. 

The centre point for calculating the curve of growth was selected to coincide
with the position of the ionising stellar cluster for the rounded, symmetrical
regions. The position of each 
cluster was determined from the observations in the  previously aligned
continuum bands.  
In the case of multiple
regions, the curve of growth starting point was found by centring the
global emission of the whole region. Although this lead to centres which
did not correspond to any observed feature in particular, the shape of the 
curve of growth was better defined and allowed a better estimation of its
asymptotic value. To do this we compared the flux increase on three
succesive rings to the estimated flux noise. When this increment was found
to be smaller than the uncertainties, the aperture integration was stopped
and the final flux value recorded.

By visual inspection of the images, the residuals due to over or under 
subtraction of stars were detected,  marked as `bad pixel' regions and
neglected when calculating the observed flux.

\subsection{Estimation of the photometric errors}

The curve of growth procedure described previously yields the total flux
detected for each region. The error bars shown in the corresponding 
plots (\ref{fig:M33plot}, \ref{fig:M101plot}, \ref{fig:NGC6822plot}) 
represent the uncertainty due to the photon counting process, 
determined by the AUTOPHOTOM package. As we are dealing with extended 
objects, the errors in the determination
of the background emission are very important. To account for this, we 
estimate the noise in the photometry from the dispersion measured in the
regions used to perform the background subtraction instead of 
using the nominal readout noise of the CCD. This ensures a
more realistic estimate of the uncertainties involved in the process.

As for the flux calibration errors,
there is an unavoidable uncertainty in the zero-point correction 
($Z_{\lambda}$) and the extinction per airmass unit ($k$). 
The overall uncertainty can be calculated then differentiating the expression
that links observed flux ($F$) and counts ($N$) in the detector:
\begin{equation}
  \label{eq:photerr}
  \Delta F = \frac{\partial F}{\partial N} \, \Delta N +
\frac{\partial F}{\partial Z} \, \Delta Z +
\frac{\partial F}{\partial k} \, \Delta k 
\end{equation}
which leads to
\begin{eqnarray}
  \label{eq:photerr2}
  \Delta F & = & 10^{\frac{Z+kX}{2.5}} \, \Delta N + 
  N \, 10^{\frac{Z+kX}{2.5}} \frac{\ln(10)}{2.5} \, \Delta Z  \nonumber \\
  &  & \mbox{}+ N \, 10^{\frac{Z+kX}{2.5}} \frac{\ln(10)}{2.5} \, X \, \Delta 
k.
\end{eqnarray}

Dividing this uncertainty by the total flux measured, we can get the 
relative error in the flux determination:
\begin{equation}
  \label{eq:relphoterr}
  \frac{\Delta F}{F} = \frac{\Delta N}{N} + \frac{\ln(10)}{2.5} \, (\Delta Z + \Delta k \, X).
\end{equation}

From  the curves of growth  (figures 
\ref{fig:M33plot} to \ref{fig:NGC6822plot}) it can be seen that the 
relative error due to the
photon counting and background noise is about 4\%, or less. The goodness
of the extinction correction was also shown in the parameters of the fit
to the standard stars included in table \ref{extincttab}. From this
table an average error of 0.04 can be estimated for the zero point correction
and of 0.03 for the uncertainty in the determination of the extinction
per airmass unit. As the airmass of the observed objects was always smaller
than 2, a rough average of 1.5 for $X$ leads to an estimate for the second term
of equation \ref{eq:relphoterr} of 7\%.
Finally a 2\% error 
should be
accounted for the `dust flakes' problem in the flat fields. Hence, we 
consider that a value of 13\% correctly represents -- conservatively -- 
the relative error in the fluxes determined in this work.

\subsection{Interstellar Extinction Correction}
\label{sec:extin}

Both $H_{\alpha}$ and $H_{\beta}$ fluxes were obtained for six of
the observed regions allowing  an extinction correction.
  We will
summarise the standard procedure  now, specifying the nomenclature used to 
avoid confusion:

The observed flux $F$ at a wavelength $\lambda$ of an intrinsic flux 
$F_0$ after passing through a 
medium of optical depth $\tau_{\lambda}$ is
\begin{equation}
F_{\lambda} = F_{\lambda 0} \, e^{-\tau_{\lambda}}.
\end{equation}

The ratio of observed to true intensities at different wavelengths is then
\begin{equation}
\frac{F_{\lambda_1}}{F_{\lambda_2}} = \frac{F_{\lambda_1 0}}{F_{\lambda_2 0}} 
\, e^{-\left(\tau_{\lambda_1} - \tau_{\lambda_2}\right)}.
\end{equation}

It is assumed that the interstellar extinction has the same 
wavelength dependence, so
\begin{equation}
\frac{F_{\lambda_1}}{F_{\lambda_2}} = \frac{F_{\lambda_1 0}}{F_{\lambda_2 0}} 
\, e^{-c\left[f\left(\lambda_1\right) - f\left(\lambda_2\right)\right]}.
\label{eq:fluxratio}
\end{equation}

Using logarithms in base 10, and 
for the case of $H_{\alpha}$ and $H_{\beta}$ we have
\begin{equation}
\frac{F_{H_{\alpha}}}{F_{H_{\beta}}} = \frac{F_{H_{\alpha} 0}}{F_{H_{\beta} 0}} 
\, 10^{-C\left[f\left(H_{\alpha}\right) - f\left(H_{\beta}\right)\right]},
\end{equation}
where $ C = 0.434\,c$, and taking logarithms
\begin{equation}
\log\left(\frac{F_{H_{\alpha}}}{F_{H_{\beta}}}\right) = \log\left(\frac{F_{H_{\alpha} 0}}{F_{H_{\beta} 0}}\right) 
 -C\,\left(f\left(H_{\alpha}\right) - f\left(H_{\beta}\right)\right),
\label{eq:extinct}
\end{equation}
allows us to calculate the extinction factor $C$. The measured ratio of
observed fluxes can be compared with the theoretical ratio of the same pair
of lines for an HII region. Case B recombination theory for an ionisation 
bound nebula at
$10^4$ K, predicts the logarithm of the ratio of $H_{\beta}$ and $H_{\alpha}$ 
fluxes to be 0.454 \cite{1971MNRAS.153..471B}. The 
$f\left(\lambda\right)$ values are obtained from Osterbrock
\shortcite{1989agna.book.....O}. This value
of $C$ is found in the literature as $C(H_{\beta})$, the logarithmic 
$H_{\beta}$ absorption coefficient.

The logarithmic $H_{\alpha}$ absorption coefficient, which we need, can be 
calculated from $C(H_{\beta})$ as
\begin{equation}
C(H_{\lambda}) = \log\left(F_{\lambda} 0\right) - \log\left(F_{\lambda}\right)
\label{eq:celambda}
\end{equation}
so, we can rearrange Eq.\ref{eq:extinct}
\begin{equation}
\log\left(F_{H_{\alpha}}\right) - \log\left(F_{H_{\alpha 0}}\right) =
-C(H_{\beta}) + C(H_{\beta})\,f,
\end{equation}
where $f$ is $f\left(H_{\alpha}\right) - f\left(H_{\beta}\right)$.
Replacing Eq.\ref{eq:celambda} for $H_{\alpha}$:

\begin{equation}
C(H_{\alpha}) = C(H_{\beta}) \, (1 -f)
\end{equation}

\section{Results}
\label{sec:photresults}

The total flux for each region was derived from the asymptotic value
determined by the curve of growth method. In order to analyse the effects 
of residuals in the sky and  continuum subtraction, we have performed
the photometry using different methods for sky estimation and masking of
the features not perfectly removed by the continuum subtraction.

The measured $H\alpha$ and $H\beta$  fluxes in the available 
bandwidths are listed in Table~\ref{obsfluxes}, together with our 
determinations for the logarithmic $H_{\alpha}$ extinction coefficient.
`Nil' values correspond to those cases in which
the measured ratio was equal to, or slightly 
lower than, the theoretical one. This might
be suggesting there is some systematic problem between fluxes determined
for both sets of filters, and we will discuss this in the following sections.

\begin{table}
\begin{center}
\begin{tabular}{lccc}
\hline\hline
GEHR id.\ & $F_{H \alpha}$ & $F_{H \beta}$ & $C_{H\alpha}$ \\
\hline
NGC~588 & -11.48 & -11.87 & 0.00 \\
NGC~592 & -11.57 & \ldots & \ldots \\
NGC~595 & -10.97 & -11.44 & 0.01 \\
NGC~604 & -10.47 & -10.93 & 0.00 \\
NGC~5447 & -11.33 & \ldots & \ldots \\
NGC~5461 & -11.19 & \ldots  & \ldots \\
NGC~5462 & -11.37 & \ldots  & \ldots \\
NGC~5471 & -11.21 & \ldots & \ldots \\
NGC~6822 I & -11.36 & -11.86 & 0.03 \\
NGC~6822 III & -11.32 & -11.87 & 0.06 \\
\hline
\end{tabular}
\caption[Results from the photometric analysis of the observed giant HII regions.]{Results from the photometric analysis of the observed giant HII 
regions. For each region, we have listed the $H\alpha$ flux in column 2,
the $H\beta$ flux in column 3, and the logarithmic absorption coefficient
in column 4.}
\label{obsfluxes}
\end{center}
\end{table}

\subsection{Comparison with previous photometric studies}

The largest collection of total fluxes of GEHRs in the Local
Group in the $H\beta$ bandwidth have been obtained by  M79 and MMTG\@.
These are summarised in Table~\ref{pubflux}. We have included in column 2,  
the expected flux of H$_{\beta}$ from K79 H$_{\alpha}$ photometry, 
calculated using the theoretical $H\alpha / H\beta$
ratio for Case-2 recombination (Osterbrock 1989). The 
discrepancies among the measured fluxes are easily noticed.
There are several possible sources for these discrepancies.
Firstly, the use of non-linear detectors can fail to register properly the
large dynamical range of fluxes in GEHRs from the weak diffuse
emission present over a large area to the strong flux emitted by
conspicuous `knots'. Secondly, the flux calibration process for narrowband
emission line photometry lacks of standard objects and a proper reference
system, such as Johnson \& Morgan \shortcite{1953ApJ...117..313J} 
for the UBV photometry. Thirdly, the 
extinction correction also varies among different studies, and can account 
for some of the discrepancies in the net fluxes calculated.

\begin{table}
\begin{center}
\begin{tabular}{lcccc}
\hline\hline
GEHR id.\ & \multicolumn{2}{c}{Kennicutt(1979)} & Melnick(1979) & MMTG \\
   &  $F_{H\alpha}$ & $F_{H\beta}^{\ast}$ & $F_{H\beta}$ & $F_{H\beta}$  \\
\hline
NGC~588      & -11.63 & -12.08 & -12.09 & -11.73 \\
NGC~592      & -11.82 & -12.27 & \ldots & -11.61 \\
NGC~595      & -11.06 & -11.51 & -11.74 & -11.58 \\
NGC~604      & -10.60 & -10.95 & -11.15 & -11.17 \\
NGC~5447     & -11.80 & -12.25 & \ldots & -12.57 \\
NGC~5461     & -11.50 & -11.95 & -11.89 & -12.29 \\
NGC~5462     & -11.82 & -12.27 & -12.37 & -12.21 \\
NGC~5471     & -11.45 & -11.90 & -11.97 & -12.10 \\
NGC~6822 I   & -11.30 & -11.75 & -12.09 & -12.48 \\
NGC~6822 III & -11.23 & -11.68 & -12.05 & -12.49 \\
\hline
\end{tabular}
\caption[Published photometry for Giant HII Regions in the Local Group.]{Published photometry for Giant HII Regions in the Local Group. This
subset includes the regions for which we have obtained new photometric
data. The $H\beta$ fluxes for K79 are estimated from the 
H$\alpha$ flux assuming a
theoretical ratio between these two lines of 2.82, only for comparison
purposes}
\label{pubflux}
\end{center}
\end{table}

The comparison of our results with previously published photometry on these
HII regions is discussed separately in the following four sub-sections, in 
order to analyse the possible sources of disagreement.
The differences between published fluxes and the ones obtained in the present 
study are presented in table \ref{compflux}. These are calculated as 
$ \Delta_{Other} = F_{H(\beta,\alpha)}^{Ours} - F_{H(\beta,\alpha)}^{Other}$ 
including available filters for each comparison. Hence, positive differences
indicate our determined fluxes are stronger.

\begin{table}
\begin{center}
\begin{tabular}{lccc}
\hline\hline
GEHR id.\ & $\Delta_{K79}$ & $\Delta_{M79}$ & $\Delta_{MMTG}$ \\
          & $F_{H\alpha}$ & $F_{H\beta}$ & $F_{H\beta}$      \\
\hline
NGC~588      & 0.17 & 0.26   & -0.10  \\
NGC~592      & 0.22 & \ldots & \ldots \\
NGC~595      & 0.08 & 0.37   & 0.21   \\
NGC~604      & 0.09 & 0.18   & 0.20   \\
NGC~5447     & 0.43 & \ldots & \ldots \\
NGC~5461     & 0.30 & \ldots & \ldots \\
NGC~5462     & 0.43 & \ldots & \ldots \\
NGC~5471     & 0.25 & \ldots & \ldots \\
NGC~6822 I   & 0.02 & 0.29   & 0.68   \\
NGC~6822 III & 0.00 & 0.32   & 0.74   \\
\hline
\end{tabular}
\caption[Differences between the emission line fluxes of our photometry and 
published values.]{Differences in the determination of emission line fluxes 
between our photometry and values available in the literature.}
\label{compflux}
\end{center}
\end{table}

\subsection{Comparison with K79}

\begin{figure}
\centerline{
\psfig{figure=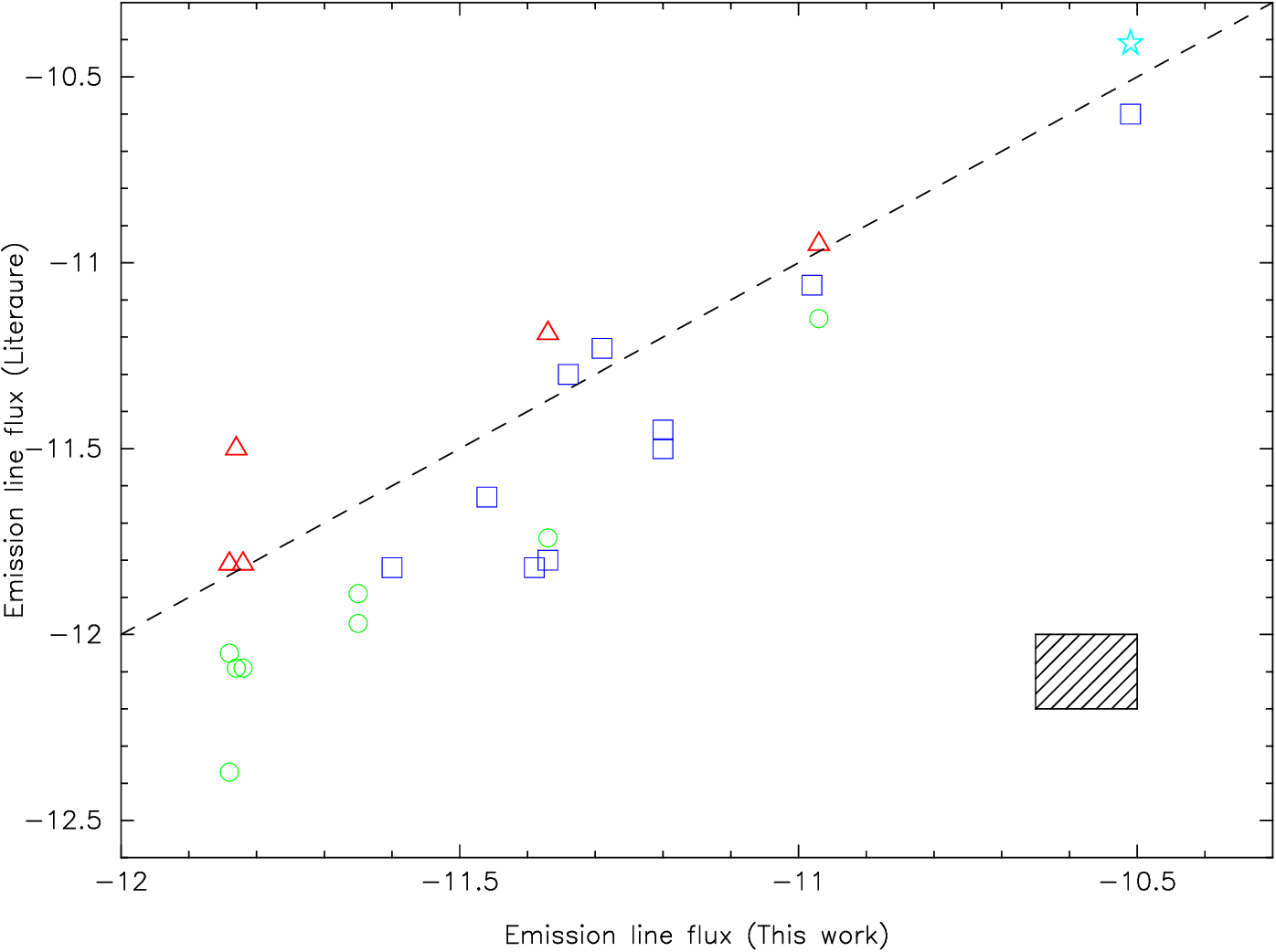,angle=270,width=8cm,height=8cm,clip=}}
\caption{Comparison between our emission line photometry and previous ones
available in the literature. Different symbols refer to different sources
for comparison as follows: squares for K79, circles for M78, triangles for
MMTG and the star symbol for Curchwell and Goss (1999). The dashed line
shows the location of identical flux determinations in the plot. A shaded box 
at the lower right corner shows an approximate range of photometric errors.}
\label{fig:fluxes_comp}
\end{figure}

K79 obtained photographic plates (on Kitt Peak's 2.1 m and 0.9 m telescopes) 
and photoelectric H$\alpha$ photometry (Kitt Peak's 0.9 m and Manastash 
Ridge 0.8 m telescopes fitted with RCA C31034 GaAs phototubes) of 
a large number of Giant HII regions which include all of our sample. 
Plotted in Figure~\ref{fig:fluxes_comp} are  our CCD 
photometry values on the x-axis {\it vs.} Kennicutt's  on the y-axis. 
Some regions show good agreement while others are too far
from the line that indicates identical values. The deviant regions  are almost
all from the same parent galaxy, M~101. Although these regions have an
integrated flux that is within the range of the other regions, they have a
much concentrated profile, with a high flux density in the inner core as seen 
from the radial flux density profiles 
(figures \ref{fig:M33plot} to \ref{fig:NGC6822plot}).
MMTG argue
that there could be some problem with saturation in Kennicutt's 
photometry, on the basis of a  comparison with their H$\beta$ photometry, 
and a low H$\alpha$ to H$\beta$ ratio.

The differences in the H$\alpha$ flux measured for regions NGC 604 and 
NGC 595 are small, but 
are slightly larger for NGC 588 and NGC 592, all of them from 
the same host galaxy, M33. It is mentioned in
K79 that photometry for lower brightness regions was obtained
from photographic plates, which would enhance non linearity problems.

Errors could also arise from any systematic but
undetected difference between the sky conditions during our observing run.
This is a possibility, as most of the 
GEHRs in M101 were observed in the second half of the run. However, the
scatter observed for the standard stars in those nights is much smaller
than the differences in flux, and although regions NGC 595 and 
NGC 5471 were observed
on the same night, only NGC 5471 shows a discrepancy between the
fluxes calculated in this work and those from Kennicutt.

\subsection{Comparison with M79}

There are five regions in common with the H$\beta$ photometry performed by
M79. In his work, the H$\beta$ emission fluxes were determined with
a two-channel photoelectric photometer attached to the 1.5m 
telescope at the Palomar Observatory. The flux calibration was performed 
through the observation of three standard subdwarf F stars. Melnick
estimates a conservative error of $\pm0.1$ mag.\ although there could be
somewhat larger systematic errors due to differences in angular sizes of
the nebulae and its effect on the illumination of the photocatode.

Figure~\ref{fig:fluxes_comp} shows a comparison between our H$\beta$ fluxes 
and the ones found in his work. The five regions in 
common are plotted with open circles. As it can be seen from the figure, 
there is a systematic difference between both results, being our values
brighter by a factor of 2. A similar systematic difference can be seen
with the H$\beta$ photometry which we will discuss below.

\subsection{Comparison with MMTG}

The comparison with MMTG photometry is more difficult, as there are five
regions in common too, and there are no details about the photometric 
determinations in their paper. In figure \ref{fig:fluxes_comp} we can
see that the overall agreement is good, although the discrepancy for
one region (NGC~588) is somewhat larger than expected. 
As it was already discussed in Section \ref{sec:photresults}, our 
H$\beta$ fluxes
might be larger than expected when compared to H$\alpha$ fluxes, 
as the ratio very rarely
exceeds the theoretical value 2.82, expected for no extinction. Furthermore
NGC~588 shows the smallest ratio (2.34) which suggests an overestimation
of our H$\beta$ flux for that region.

MMTG show in their work a very good agreement with M79 photometry, which might
seem puzzling as our data agrees only with one of them. This arises from
the fact that the agreement between the previous studies holds for the larger
sample in common, but not for the subset of the regions included in our sample.

Still, our H$\beta$ photometry is not as reliable as expected. Keeping this 
in mind, we will base most of our further analysis on the H$\alpha$ photometry.

\subsection{The case of NGC 604}

In a recent paper Churchwell \& Goss \shortcite{1999ApJ...514..188C} 
study the extinction within NGC 604 based on their own radio wavelength data 
from the Very Large Array (VLA) and CCD H$\alpha$ photometry performed
on an HST image from D.\ Hunter. We will further discuss their 
findings later in this paper. Their determined
total H$\alpha$ flux for the region, 
$(3.9 \pm 0.4) \times 10^{-11}$ ergs cm$^{-2}$ s$^{-1}$ is in agreement 
within the errors with our value  
of $(3.1 \pm 0.4) \times 10^{-11}$.

\section{Extinction maps}
\label{sec:extinction}

From the overall fluxes determined for each region we have calculated
the ratio between the H$\alpha$ and H$\beta$ fluxes, in order to
estimate the extinction affecting the measured H$\alpha$ flux. However,
it is quite well known for star forming regions that the nebular
characteristics are far from homogeneous. This applies to the
extinction properties of the interstellar medium, which combines neutral,
ionised and molecular gas, plus dust in different geometrical distributions.
Churchwell \& Goss \shortcite{1999ApJ...514..188C} have used the ratio
between radio continuum and H$\alpha$ to map the extinction properties in 
NGC~604 and analyse their correlation with the CO distribution from the
work of Viallefond et al.\ \shortcite{1992A&A...265..437S}. A low
spatial resolution map was constructed by Ma\'{\i}z Apell\'aniz 
\shortcite{1999PhD...........M} using longslit flux calibrated
spectra taken at different positions in NGC 604. 

In order to obtain a `viewable' version of the flux ratio map, the calibrated
nebular images had to be filtered to remove low signal pixels that 
introduce large scatter when performing the pixel to pixel division. For the 
sake of clarity, we have removed all pixels which had flux lower than
seven times the background noise. After performing the division, the 
spatial variations were smoothed filtering the resultant image. We chose
a circular median filter with a four pixel radius, by means of the IRAF task
{\tt rmedian}. Although this introduces a degradation in the image spatial 
resolution, it helps to visualize the main trends in extinction changes.

From our data, we can produce extinction maps for NGC~604, NGC~595, and
both observed regions in NGC~6822. These are shown in figures 
\ref{fig:ngc604ext} to \ref{fig:ngc6822IIIext}. 

\begin{figure}
\centerline{
\fbox{Figure17}}
\caption{Ratio of H$\alpha$ to H$\beta$ flux for NGC~604. The image has a 
larger scale
than the nebular filter image, although it shares the coordinates origin}
\label{fig:ngc604ext}
\end{figure}

\begin{figure}
\centerline{
\fbox{Figure18}}
\caption{Same as figure \ref{fig:ngc604ext} for NGC~595}
\label{fig:ngc595ext}
\end{figure}

\begin{figure}
\centerline{
\fbox{Figure19}}
\caption{Same as figure \ref{fig:ngc604ext} for NGC~6822~I}
\label{fig:ngc6822Iext}
\end{figure}

\begin{figure}
\centerline{
\fbox{Figure20}}
\caption{Same as figure \ref{fig:ngc604ext} for NGC~6822~III}
\label{fig:ngc6822IIIext}
\end{figure}

The extinction varies strongly throughout most of the GEHRs (the 
H$\alpha$/H$\beta$ flux ratio varies from 2.5 up to 7), 
which can be readily explained when considering
the dramatic changes in the distribution of obscuring material in these
star forming regions. 
NGC~6822~I
turns out to be a singular case, as the 
H$\alpha$/H$\beta$ flux ratio remains constant and close to the
theoretical expected value. This can be interpreted as a lack of obscuring 
matter (such as molecular clouds) which might suggest there is no material
left over for further, or ongoing, star formation. We will discuss this issue
in section \ref{sec-trueGEHRs}.

\section{The Luminosity - Velocity dispersion ($L$ - $\sigma$) relation for 
GEHRs}
\label{sec:lsigma}

\subsection{Luminosities}

With the observed fluxes obtained in the previous section it is possible
to determine the luminosity by correcting for the distance $d$ to the parent
galaxy. The correction factor is simply $4\,\pi\,d^2$, with $d$ expressed in
cm. Distances are usually expressed in distance 
modulus $(m-M)$ related by
$M = m + 5 - 5 \, \log(D)$, where $D$ is the distance in parsecs. 
The determination of accurate distance to the
galaxies of the Local Group has been the central aim of several research
projects. Distances to NGC 6822, M~33 and M~101 have recently been 
determined with
primary indicators, such as Cepheids, by the HST Extragalactic Distance Scale 
Key Project \cite{1991ApJ...372..455F}, \cite{1996ApJ...463...26K}. 
We have used the cepheid distance scale for the galaxies hosting the GEHRS
included in this study and listed in Table \ref{galaxydist}.
This indicator also agrees well with other 
determinations such as the tip of the red giant branch and the planetary nebula
luminosity function, as discussed in Kennicutt et al.\ 
\shortcite{1998ApJ...498..181K}. Note that the estimated 
distances to the host galaxies include a considerable source of uncertainty
which adds to the flux errors determined above.

\begin{table}
\begin{center}
\begin{tabular}{lcccc}
\hline\hline
Galaxy & $(m-M)_{PL}$ & $\log(4 \pi d^2)$ & $\Delta_{HH86}$ & $\Delta_{MMTG}$ \\
\hline
M~33     & 24.63 $\pm$ 0.09 & 49.93 & -0.23 & 0.18 \\
M~101    & 29.35 $\pm$ 0.17 & 51.82 & +0.26 & 0.06 \\
NGC~6822 & 23.62 $\pm$ 0.20 & 49.53 & +0.09 & 0.13 \\
\hline
\end{tabular}
\caption[Distance to host galaxies.]{Distance to the galaxies hosting the 
GEHRs included in this study.}
\label{galaxydist}
\end{center}
\end{table}

In order to compare with previous determinations, we have included results from
M79, 
Hippelein \shortcite{1986A&A...160..374H} (HH86), and MMTG. However, it must be
kept in mind that each study has made use of the available distances from
different sources and discrepancies in these distances introduce different
corrections. We have included in Table \ref{galaxydist} the correction to the
logarithmic luminosity calculated with the distances adopted in this work, 
and the 
differences with the same factors derived for the distances used in previous 
works. It can be readily seen  that the
distance to the galaxy is
a considerable source for disagreement in luminosities. The magnitude of
these disagreements on the distance scales used and the problems that may have
affected previous photometric studies, as described in the previous 
section make any comparison of the calculated luminosities almost impossible.

\begin{table}
\begin{center}
\begin{tabular}{lcccc}
\hline\hline
 & M79 & 
   HH86 &
   MMTG &
   {This work} \\
GEHR id.\ & $I_{H\beta}$ & $I_{H\alpha}$ & $I_{H\beta}$ & $I_{H\alpha}$ \\
\hline
NGC~588     & 38.08  & 38.61 & 38.25 &  38.47  \\
NGC~592     & \ldots & 38.42 & 38.44 &  38.33  \\
NGC~595     & 38.64  & 39.18 & 38.55 &  38.95  \\
NGC~604     & 38.83  & 39.64 & 38.80 &  39.42  \\
NGC~5447    & \ldots & 39.82 & 39.59 &  40.45  \\
NGC~5461    & 39.93  & 40.12 & 39.77 &  40.62  \\
NGC~5462    & 39.72  & 39.80 & 39.65 &  40.43  \\
NGC~5471    & 39.73  & 40.17 & 39.85 &  40.62  \\
NGC~6822I   & 38.46  & 38.36 & 37.59 &  38.21  \\
NGC~6822III & 38.30  & 38.44 & 37.59 &  38.30  \\
\hline\hline
\end{tabular}
\caption[GEHRs luminosities.]{Luminosities 
available in the literature, plus those calculated in 
this paper.}
\label{absphotom}
\end{center}
\end{table}

\subsection{Velocity dispersions}

Velocity dispersions have been independently obtained by several groups
analysing the kinematics of GEHRs. These efforts can be split 
according to the spatial resolution of the instruments 
used for the estimations into large and small scale observations. I will 
discuss in this section the large scale observations, as these are directly
linked to the overall luminosity measured for the HII regions. Melnick 
\shortcite{1977ApJ...213...15M},  measured line profiles and determined 
turbulent velocities for a set of GEHRs in nearby galaxies 
using a Fabry-Perot interferometer at Palomar observatory
when investigating the relation of the velocity dispersion with the linear
diameter of giant HII regions. The turbulent velocity $\sigma$ is
determined from the observed line profile $\sigma_o$ after correcting
from the instrumental profile $\sigma_i$ and the thermal contribution to
the observed width of the emission line $\sigma_T$. Hence:
\begin{equation}
\sigma^2 =  \sigma_o^2 - \sigma_i^2 - \sigma_T^2
\end{equation}
HH86 obtained newer F-P interferometric 
data, including a larger sample of regions with
dissimilar results. Independently Roy et al.~\shortcite{1986ApJ...300..624R} 
(RAJ86) obtained another set of velocity dispersions for
extragalactic HII regions. The agreement of both sets of measurements for the
regions included in our study can be easily seen in Table \ref{dispvelocities}.
In a later analysis MMTG combined the
data available for H$\alpha$ and [OIII] from HH86 and characterised the
gas kinematics with $\sigma = \frac{\sigma_{H\alpha} + \sigma_{[OIII]}}{2}$.

\begin{table}
\begin{center}
\begin{tabular}{lccc}
\hline\hline
GEHR id.\ & HH86 & RAJ86 & MMTG \\
\hline
NGC~588     & 12.9 $\pm$ 0.4 & 14.9 (21.1 $\pm$ 1.0) & 12.8 \\
NGC~592     & 13.6 $\pm$ 1.0 & 12.1 (17.1 $\pm$ 1.0) & 12.6 \\
NGC~595     & 17.8 $\pm$ 0.8 & 19.2 (27.1 $\pm$ 1.6) & 17.2 \\
NGC~604     & 17.8 $\pm$ 0.3 & 16.3 (23.1 $\pm$ 0.8) & 16.8 \\
NGC~5447    & 20.3 $\pm$ 0.6 & 21.2 (30.0 $\pm$ 1.2) & 19.3 \\
NGC~5461    & 21.5 $\pm$ 0.3 & 24.7 (35.0 $\pm$ 1.3) & 20.3 \\
NGC~5462    & 24.2 $\pm$ 0.4 & \ldots                & 21.8 \\
NGC~5471    & 22.4 $\pm$ 0.4 & 21.3 (30.2 $\pm$ 1.0) & 21.0 \\
NGC~6822I   & 10.5 $\pm$ 0.2 & 9.5  (13.4 $\pm$ 0.8) & 9.8  \\
NGC~6822III & 10.3 $\pm$ 0.3 & 10.4 (14.7 $\pm$ 0.9) & 9.5  \\
\hline
\end{tabular}
\caption[GEHRs velocity dispersions of the ionised gas.]{Velocity dispersions 
from the literature, of the ionised gas  
of the GEHRs included in this study. We have listed
the values of the $\sigma$ dispersions published, with the exception of the
values for RAJ86, which we scaled from the beta width listed in their paper,
included between parentheses in column 3.}
\label{dispvelocities}
\end{center}
\end{table}

There are several other studies of spatial variations in the
kinematics of these large regions. These were done mostly with ech\'elle 
spectroscopy (Skillman \& Balick \shortcite{1984ApJ...280..580S}, 
Chu \& Kennicutt \shortcite{1994ApJ...425..720C}) and with the TAURUS 
Fabry-Perot Imaging Spectrograph on the 4 m.\ William Herschel Telescope in 
La Palma (Mu\~noz-Tu\~n\'on et al.~\shortcite{1995AJ....110.1630M}, 
\shortcite{1996AJ....112.1636M}, Yang et al.~\shortcite{1996AJ....112..146Y}).
TAURUS \cite{1980MNRAS.191..675T} allows to obtain simultaneously 
high resolution spectroscopic information with its etalon grating and seeing
limited imaging.
These studies added new information regarding the smaller scale kinematics
of these regions, showing there is a combination of shell structures, loops
and supersonic motions down to the smallest observable detail. We will discuss 
these important results in section \ref{sec-trueGEHRs}.

\subsection{The $\log(L(H\alpha))$ vs $\log(\sigma)$ regression}

As already mentioned, several studies analysed the existence of a
correlation between the total luminosity emitted by a giant HII region in one
of its Balmer lines and the velocity dispersion of the gas measured from
the width of such line. The slope derived for this
regression is  a crucial parameter to find an explanation for the
source of such a relation. In this section we will analyse and discuss
the existence of this correlation on the basis of our photometry.

We have combined the luminosities obtained in the previous section from
our CCD photometry and the kinematical data used in MMTG
 in order to obtain an improved set of data for analysing their relation. 
The existence of a correlation was analysed by performing a least squares 
fit to the
available data, accounting for errors in both coordinates. Errors were
estimated to be about 10\% in the kinematics and 20-25\% in the luminosities. 
In the logarithmic plane, a linear regression is found, where 
\begin{equation}
  \label{allreg}
  \log(F(H\alpha))=29.6 (\pm 0.7) + 8.2 (\pm 0.6) \times \log(\sigma). 
\end{equation}

\begin{figure}
\centerline{
\psfig{figure=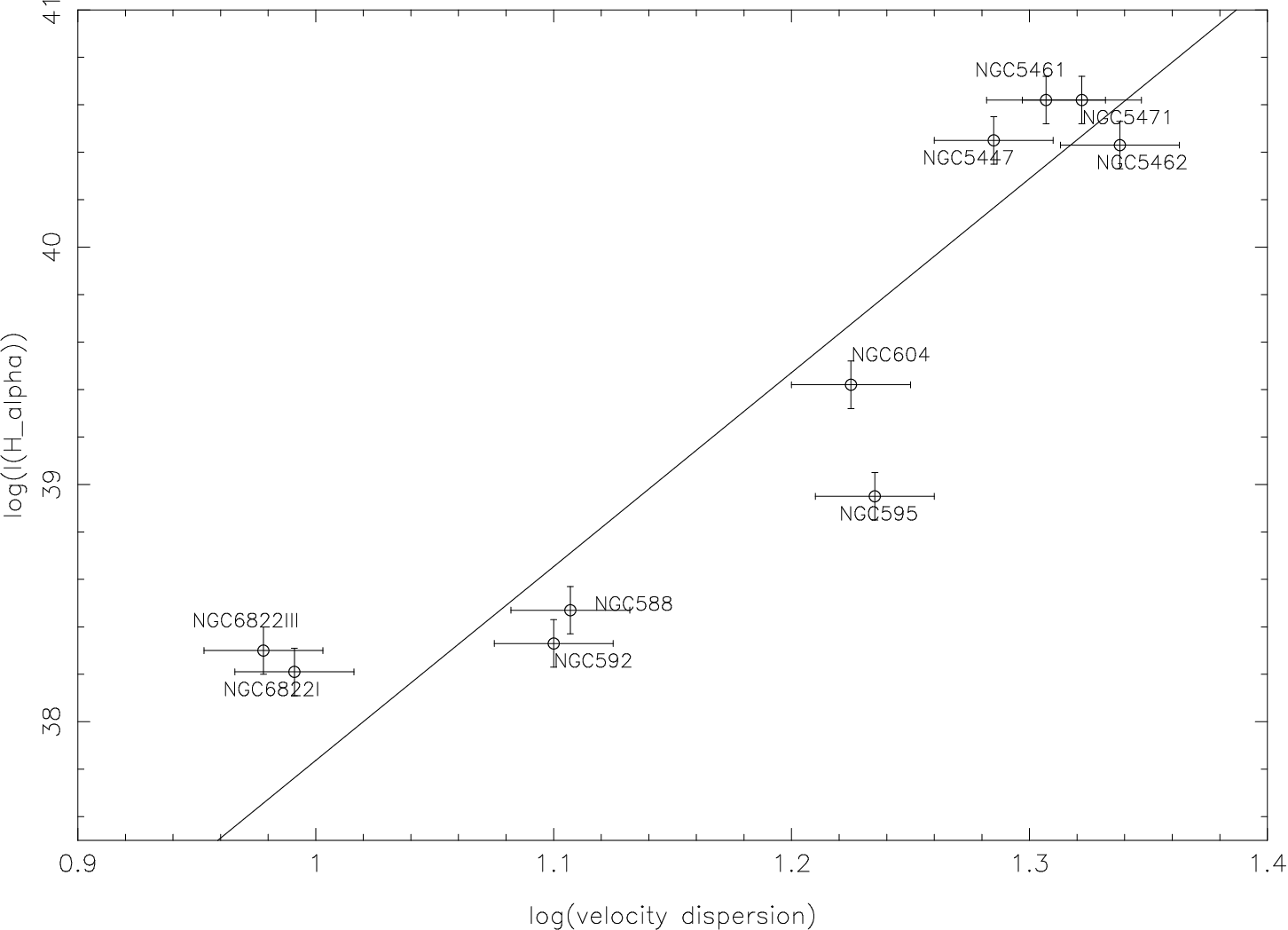,angle=270,width=8cm,height=8cm,clip=}}
\caption[H$\alpha$ luminosity vs.\ $\log(\sigma)$.]{Plot of the logarithm of the H$\alpha$ luminosity against the
logarithmic velocity dispersion. A linear fit to the data is also drawn.}
\label{Ivss_all}
\end{figure}

The data is plotted in Figure \ref{Ivss_all} together with the line
showing the fit to the points. From the plot it can be readily seen that
although there is a correlation, there is also a large scatter 
in the data.
Deviations from the fit are too large for the expected uncertainties 
(a $\chi^2$ test on the fit yields a probability almost zero) and
large enough to seriously limit its use to derive accurate
luminosities from measured emission line widths.

It seems that, under the evidence of the new photometry here presented, the
uncertainties in previous determinations of giant HII region luminosities 
were larger than estimated. Also, there is increasing evidence that there
is an age spread between GEHRs and that this could modify
noticeably the observed global parameters.
However, our knowledge of the finer details of some of these
regions can help us select a more homogeneous sample.

\subsection{The young  GEHRs}
\label{sec-trueGEHRs}

The small scale kinematic studies brought new information to the
analysis of GEHRs. Most of them showed that the supersonic 
profile obtained with global spectroscopic data is formed by the 
contribution of different regions within each GEHR with different kinematic
behaviour. It was also found that this mixture of different contributions
was not uniform for all HII regions; some regions have most of their flux
coming from supersonic cores, others have a larger contribution from bubbles
and shells. We have used this information to select a more homogeneous set
of GEHRs out of our sample. Particularly for the farthest regions, for which
the angular resolution makes any guessing of its structure impossible. 
Based on the assumption that the large shells and bubbles are formed at 
later stages of the evolution of these star forming regions is that we
will refer to those regions that do not show evidence of loops and filaments 
as the \em young \em GEHRs. We will use spatially resolved kinematical
data to disentangle the multiple blobs that form the M~101 GEHRs and
identify those who show supersonic velocity widths.
Including the available information, and making use of the excellent seeing 
conditions
for our photometric data we have re-analysed all the regions 
in our JKT photometry sample. 

\subsubsection{M~33 GEHRs}

The galaxy M~33 is situated at about 840 kpc \cite{1991ApJ...372..455F}. 
The whole emission line region
of the 30 Doradus nebula in the Large Magellanic Cloud, which is the one for 
which we have the best detailed
information, would occupy only 1 arcmin, instead of the 15 arcmin it fills
at its actual distance. However, at this distance it is still possible to 
resolve the
presence of large bubbles, shells or any other evidence of nebular disruption
and combine it with other information regarding its age or evolutionary
stage of its components.

\vskip 0.2cm
\noindent{\bf NGC~588}

This region shows clear evidence, as seen in figure \ref{fig:ngc588}
that there is a compact core, but it also has at least two strong ring-like
structures that contribute to the emitted flux. Kinematic data available from
MTCT indicate the presence of 
shells with intensity comparable to
the rest of the kinematic core. From the expected traces of evolution
of GEHRs we consider this region as an older region
which has been swept by strong winds and supernova explosions.

\vskip 0.2cm
\noindent{\bf NGC~592}

The H$\alpha$ image shows two nuclei dominating the emission of the core of
the region. A filamentary nebula surrounding the core and extending out to
large distances, makes it difficult to find an asymptotic
value to determine the total flux (see Figure \ref{fig:ngc592}).
There is no data about the small scale kinematics of this region, 
although its structure reveals a strong interaction with the external
environment, suggesting an older age too. The presence of two condensed
knots makes it an interesting target for analysing their kinematical
properties with TAURUS-like Fabry-Perot interferometers.

\vskip 0.2cm
\noindent{\bf NGC~595}

This region has a structure quite similar to that of NGC~592. There are a
number of dispersed strong knots, visible in figure \ref{fig:ngc595}. 
There is no data
available from small scale kinematics for this region. However, Malumuth
et al.\ \shortcite{1996AJ....111.1128M} have obtained multiband HST/WFPC2
images of the ionizing stellar cluster.
They derive an age of 4.5 Myr which indicates it is a somewhat evolved
cluster consistent with the presence of Wolf-Rayet stars.
\cite{1993AJ....105.1400D}.

\vskip 0.2cm
\noindent{\bf NGC~604}

The region NGC~ 604 shows an extended structure, and some small loops can 
be identified within the nebula. There is, however no strong evidence of
disruption of the nebula, as the stellar cluster remains enshrouded by the
nebula in the H$\alpha$ image shown in figure \ref{fig:ngc604}. A comprehensive
investigation of the kinematics of NGC~ 604 was carried out by Yang et al.\
\shortcite{1996AJ....112..146Y}. In their work they gathered longslit echelle
spectroscopy and TAURUS data and detected  expanding shells
together with an overall presence of broad profiles in every position within
the region. They plotted the measured FWHM of the profiles versus their 
intensity and found that the brightest profiles define a narrow band in such
diagram. This band, with a median value of $\sim$ 36 km s$^{-1}$ include most
profiles of the distribution, and dominates the total observed flux. 
TAURUS data in MTCT shows a 
similar distribution in the $I$ vs.\ 
$\sigma$ diagram. It was proposed in both papers that gravitation would be
responsible for the profiles in the narrow band, and hence dominate the
kinematics of the region. 
Gonz\'alez Delgado \& P\'erez \shortcite{astro-ph/0003067} combined archival 
HST data, together with spectra from the International Ultraviolet 
Explorer (IUE) and optical spectra from the William Herschel Telescope (WHT)
to analyse the evolutionary state of the region. They combined the
available information from stellar wind resonance lines, nebular optical
emission lines and HeI absorption lines present in the spectra of NGC~604 to
derive a 3 Myr age for a single burst that originated the central ionising 
cluster.
All these facts indicate that NGC~604 can be considered as a young GEHR.

\subsubsection{M~101 GEHRs}

The spiral galaxy M~101 lies at about 7.4 Mpc from us 
\cite{1996ApJ...463...26K}. This is ten times more
distant than M~33 and consequently with a linear resolution one order of
magnitude smaller. The apparent size of the 30~Dor nebula at this distance
would be of only
6 arcsecs. It is very difficult to base an analysis simply on the
distribution of the emission flux, but we will rely on the information
available from other kinematic studies, whenever available.

\vskip 0.2cm
\noindent{\bf NGC~5447}

This region, as almost all regions in M~101 has already been classified
as `multiple' due to the presence of several bright components 
\cite{1982ApJ...255L..29L}. We have included an expanded view of this region
in figure \ref{knots5447} and labelled the most conspicuous of these
components or `knots'. Of
particular interest is knot A, which is the strongest of all. There is
no detailed kinematics for this region, although HH86 took spectra of the
northern and southern part and found them to be almost identical. Considering
this information we performed aperture photometry, taking increasing apertures
centred in NGC~5447 A, using the local diffuse nebula to estimate the
background level. In this way the net H$\alpha$ luminosity of this particular 
knot is
calculated, and found to be $\sim 7.5 \times 10^{-13}$ ergs s$^{-1}$.

\begin{figure}
\centerline{
\fbox{Figure22}}
\caption[Expanded view of the components of NGC~5447.]{Expanded view of the 
components of NGC~5447. The most important 
knots are marked and labelled. The circle drawn illustrates the
apparent size that the 30~Dor nebula would have at the distance of M~101.}
\label{knots5447}
\end{figure}

\vskip 0.2cm
\noindent{\bf NGC~5461}

The multiplicity of NGC~5461  had already been suggested by
Kennicutt \shortcite{1984ApJ...287..116K} and analysed by MGC95 with TAURUS.
The knots, as identified by MGC95 are marked on our H$\alpha$ image in
figure \ref{knots5461}. It can be predicted from the image that knot A 
dominates
the overall flux. TAURUS data revealed that `well-behaved' Gaussian emission
lines show a velocity dispersion of 25 km s$^{-1}$. Furthermore, these
supersonic sigma values are restricted to knots A and B only. Knots C and E
are kinematically detached from the region. This can be 
guessed from the figure, although the definitive confirmation is given by the
resolution available with 2D spectroscopy.

In a similar way as described for NGC~5447 A we have calculated the H$\alpha$
luminosity emitted by NGC~5461 A and found it to be $\sim 2.6 \times 10^{-12}$
ergs s$^{-1}$.

\begin{figure}
\centerline{
\fbox{Figure23}}
\caption[Expanded view of the components of NGC~5461.]{Same as fig \ref{knots5447} for NGC~5461. The nomenclature used for the knots follows
that of Mu\~noz-Tu\~n\'on \shortcite{1994vsfd.conf.....M}.}
\label{knots5461}
\end{figure}

\vskip 0.2cm
\noindent{\bf NGC~5462}

The complex structure of this giant HII region is shown in figure 
\ref{knots5462}. Two major knots, labelled A and B are much brighter than
the remaining knots, but this particular region does not show a single
dominant feature as is the case of the previous ones.
There is no information available regarding the kinematical behaviour
of these knots as to assign the characteristics of the global profile
to any of these knots in particular. The approximate luminosities of knots 
A and B emitted in H$\alpha$ are 3.5 
and 4.5 $\times 10^{-13}$ ergs s$^{-1}$ respectively.

\begin{figure}
\centerline{
\fbox{Figure24}}
\vspace{1cm}
\caption[Expanded view of the components of NGC~5462.]{Same as fig \ref{knots5447} for NGC~5462.}
\label{knots5462}
\end{figure}

\vskip 0.2cm
\noindent{\bf NGC~5471}

NGC~5471 also shows multiple components. Skillman 
\shortcite{1985ApJ...290..449S}
analysed spatial variations for this region and identified 5 knots which
are labelled in figure \ref{knots5471}. In that figure it can also be seen
another weak component, A', detected under better seeing conditions by
MGC95. This latter study suggests that a large
fraction (80\%) of the region could be characterized by a constant emission
linewidth of $\sigma \sim 20$ km s$^{-1}$. We have estimated the 
H$\alpha$ luminosity emitted by knot A to be $ \sim 10^{-12}$ ergs s$^{-1}$.

\begin{figure}
\centerline{
\fbox{Figure25}}
\caption[Expanded view of the components of NGC~5471.]{Same as fig 
\ref{knots5447} for NGC~5471. Knots are labelled
according to  Skillman \shortcite{1985ApJ...290..449S}.}
\label{knots5471}
\end{figure}

\subsubsection{NGC~6822 GEHRs}

NGC~6822 is an irregular galaxy, situated at approximately 530kpc
\cite{1998ApJ...498..181K}. It is the
closest one from the sample studied here, and therefore the one that allows
the best linear resolution, at 2.5 pc arcsec$^{-1}$. O'Dell, Hodge and 
Kennicutt \shortcite{1999PASP..111.1382O} obtained HST photometry (using 
mainly emission line filters) and determined for both stellar clusters
associated with the two regions analysed in this paper, an
approximate age of 4 Myr.

\vskip 0.2cm
\noindent{\bf NGC~6822 I}

This region (figure \ref{fig:ngc6822I}) shows a spherical concentrated bulk, 
and an extended loop (or pair of loops). There is no
detailed kinematical data, but one can expect that the global emission
profile is dominated by the central core. However, the extinction map
reveals that there is almost no dust left. This behaviour is not observed 
in other regions that show recent star formation activity and might
suggest some evolution effects, though not as strong as in other GEHRs.

\vskip 0.2cm
\noindent{\bf NGC~6822 III}

This region has the emission highly concentrated too, with a bright nucleus
and a nebular envelope, with no evidence of shell-like features (see
figure \ref{fig:ngc6822III}).

\subsubsection{The $\log(L(H\alpha))$ vs.\ $\log(\sigma)$ for the young
and single GEHRs}

The analysis indicates that the GEHRs included in this study do not
necessarily share the same global properties. This is evident in the
case of total luminosity and global velocity dispersion. From the
expected evolution of the kinematics of giant HII regions in the frame
presented by Tenorio Tagle et al.\ \shortcite{1993ApJ...418..767T} and
MTCT 
\em young \em GEHRs (those that do not show signs of evolution) should
be governed by the gravitational supersonic motion induced by the stirring 
stars. To test this hypothesis, we have selected a sub sample that includes 
only those regions which do not show evident signs of evolution, such as
loops or shells. We have also
included the dominant cores of the multiple regions in M~101, as their
sizes are comparable to the 30~Dor Nebula, suggesting by analogy, that these 
cores are the most prominent members of an agglomeration of HII regions. 

The selection process is  summarised in Table \ref{beautycontest}.
Although the age determinations are very dependent on the used 
evolutionary tracks for each case, we are discarding regions that are 
4.5 Myr old, or
seem to be even older. Regarding the multiplicity, we are only including knots
that have properties that suggest the existence of a kinematical core.

\begin{table}
\begin{center}
\begin{tabular}{lcc}
\hline\hline
GEHR name & Age & Kinematical core \\
\hline
NGC~588 & $>4.5$ Myr & \ldots \\
NGC~592 & $>4.5$ Myr & \ldots \\
NGC~595 & $4.5$ Myr  & \ldots \\
NGC~604 & $3$ Myr    & \ldots \\
NGC~6822~I & 4 Myr   & \ldots \\
NGC~6822~III & 4 Myr   & \ldots \\
NGC~5447 & \ldots & NGC~5447A \\
NGC~5461 & \ldots & NGC~5461A \\
NGC~5462 & \ldots & nil     \\
NGC~5471 & \ldots & NGC~5471A \\
\hline\hline
\end{tabular}
\end{center}
\caption{Summary of analysis performed over the GEHRs sample to detect
regions with signs of evolution or multiplicity. Ages are as derived from
studies of the stellar clusters associated with the regions, or from an assumed
evolution of the nebular features (see text). The identification of 
kinematical cores was made from spatially resolved kinematical data of the
multiple regions in M~101.}
\label{beautycontest}
\end{table}

\begin{figure}
\centerline{
\psfig{figure=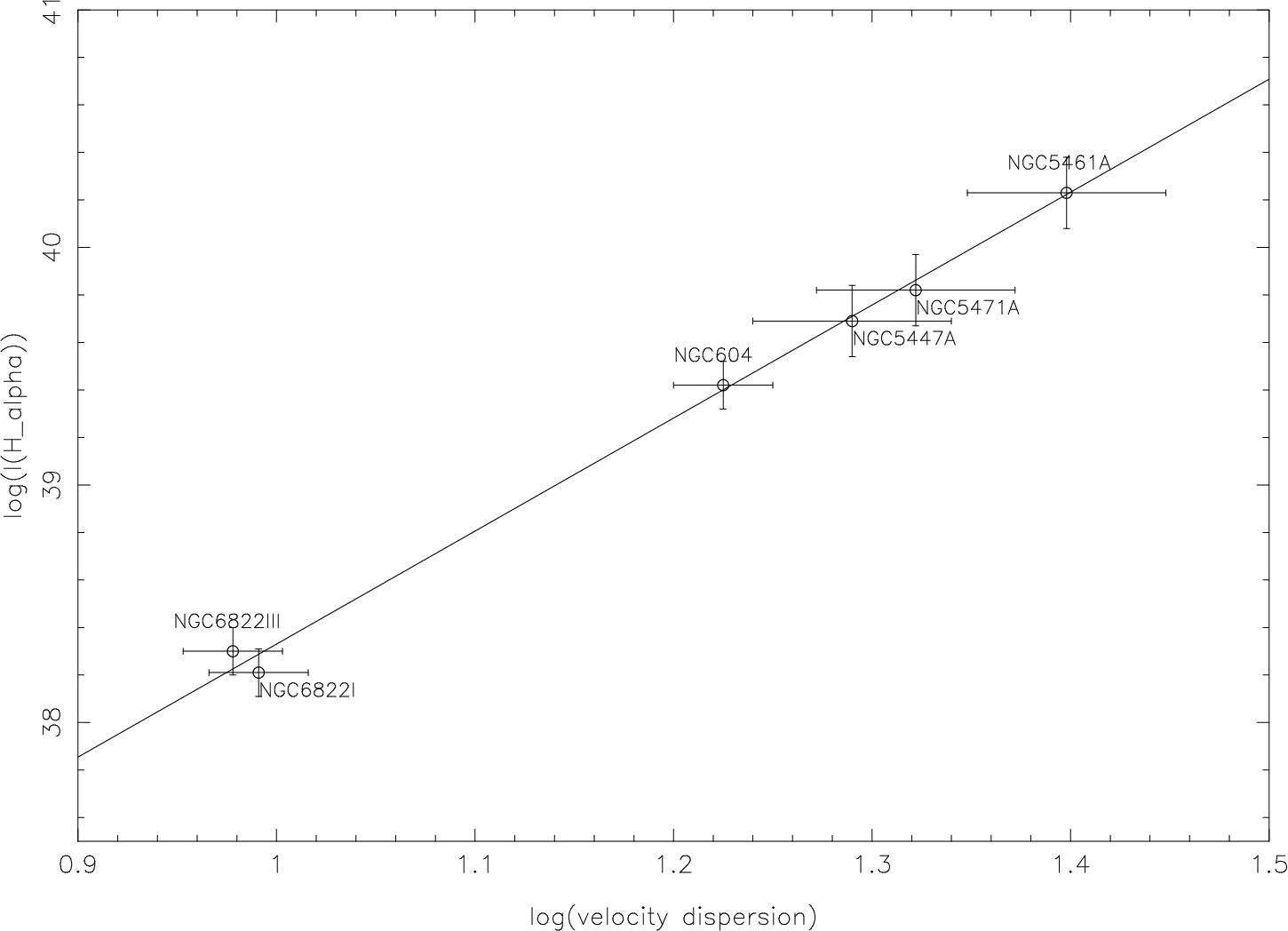,angle=270,width=8cm,height=8cm,clip=}}
\caption[Plot of the logarithm of the H$\alpha$ luminosity against the
logarithmic velocity dispersion for a subsample of GEHRs.]{The 
logarithm of the H$\alpha$ luminosity against the
logarithmic velocity dispersion for the subsample described in this section.
The regression line with slope $4.73 $ is also shown.}
\label{Ivss_good}
\end{figure}

No matter what quantitative criteria is established, the final decision 
turns out to be subjective at this stage. There is a compromise between 
discarding any region that shows signs of evolution and keeping a fair
number of GEHRs as to attempt a regression analysis.
Therefore, the short list includes: NGC~6822 I, NGC~6822 III, NGC~604, 
NGC~5471 A, NGC~5461 A,
and NGC~5447 A. A similar plot as the one presented in figure \ref{Ivss_all}
but for the mentioned subsample is shown in figure \ref{Ivss_good}. 
In it we have included the
linear fit to the points, which was found to be
\begin{equation}
  \label{youngreg}
  \log(F(H\alpha))=33.6 (\pm 0.6) + 4.7 (\pm 0.9) \times \log(\sigma).   
\end{equation}
The slope and zeropoint are -- within the errors -- in agreement with the 
values found by Fuentes-Masip et al.\ \shortcite{2000AJ....120..752F} for
their sample of high surface brightness giant HII regions in NGC 4449,
supporting a virialised system as the origin of the supersonic motions.

\section{Discussion}
\label{photdiscussion}

This paper presents the results of new $H\alpha$ and $H\beta$ CCD photometry of
ten Giant HII Regions lying in  Local Group galaxies. The 
comparison of these fluxes with previous work suggests some
systematic errors affecting the flux determinations made with photoelectric
detectors. 

A serious problem is the lack of a uniform method for
flux calibration and for that we have introduced
the use of spectrophotometric standards in order to rigorously
calculate the continuum flux correction. Accounting for the fact that the 
filter used is broader
than the emission feature is very important, and it
is generally not discussed in papers dealing with this issue. Accurate
knowledge of the emission line characteristics are crucial in order to 
calculate its position within the filter transmission curve and its
relative width. All these make the narrowband photometry of emission line
objects a very difficult one, specially when they are extended objects and
aperture spectrophotometry is not possible either.

The analysis presented here has two important although apparently 
dissimilar results:

\begin{itemize}
\item Giant Extragalactic HII Regions are far too heterogeneous objects as
to blindly include them in a sample for use as an accurate (rms $< 0.2 \log(L(H\alpha)$) distance indicator.
\item Young GEHRs and the kinematical cores of GEHRs included in this study do
follow a tight relation in the $L(H\alpha)$ vs.\ $\sigma$  plane. 
\end{itemize}
 
The first item does not preclude the use of GEHRs as a distance indicator, but 
establishes that a careful selection based on the kinematical
properties -- or ages -- of these is a must, as the
luminosities of the young GEHRs and kinematical cores seem to correlate 
very well with the velocity dispersion of the region. 
The use of the total H$\alpha$ (or H$\beta$) luminosity of the
region has not eliminated the problem of dealing with other parameters of the 
nebula, as it was hoped it would. What were
thought to be `supergiant' GEHRs (such as NGC~5461 and NGC~5471) should
be better considered as a group of HII regions, out of which one (or several) 
of them might be a true GEHR. The emission line flux originated in non-giant
HII regions can be partially responsible for the observed scatter in the
regression.

The second item mentioned above indicates that another factor 
that turns out to be very important
is the age, or evolutionary stage, of the regions. It has already been shown
for the particular case of 30 Doradus \cite{1999A&A...347..532S}, that older 
bursts of 
star formation can coexist in an active star forming region. Hence, the
presence of clear signatures of recent massive star formation do not
necessarily indicate that we are looking at an extremely
young HII region. Although
the most recent burst of stars will be responsible for the bulk
of the ionisation of the gas -- and hence of the emitted flux -- the effect
of previous stellar generations could have modified substantially the
interstellar gas kinematics.

New observational efforts should be directed to the use of 
imaging spectrographs to analyse Giant HII Regions
in the Local Group to learn more about  the kinematical 
characteristics of the emitted flux. Other global properties from HII
regions, such as the equivalent width $W_{H\beta}$ of the H$\beta$
line known to vary with the evolutionary stage
(Dottori~\shortcite{1981Ap&SS..80..267D}, 
Copetti et al.~\shortcite{1986A&A...156..111C}) 
can be used to estimate the age of the GEHRs. 
A calibration of a larger sample will be decisive to
prove what is suggested from the data now available and to provide
a link with what is observed for starburst galaxies.

\section*{Acknowledgments}

The authors wish to thank fruitful discussions with Casiana Mu\~noz-Tu\~n\'on
and Enrique P\'erez.
The JKT is operated on the island of La Palma by the Isaac Newton Group in 
the Spanish Observatorio del Roque de los Muchachos of the Instituto de 
Astrof\'\i sica de Canarias.
We acknowledge PPARC funding for the observing run at La Palma, where the ING
staff has been more than helpful. Special thanks again to Don Pollaco for 
lending us his own narrowband filter for H$\beta$. Much of this work was
developed during visits to  INAOE who hosted and supported our stays there.
GLB thanks partial funding from UNESCO and the Universidad Nacional de La
Plata (UNLP) that supported travelling expenses.
The authors acknowledge the data analysis facilities provided by the Starlink 
Project which is run by CCLRC on behalf of PPARC. In addition, the following
Starlink packages have been used: PHOTOM, GAIA.

\end{document}